\documentclass[journal]{IEEEtran}

\usepackage[latin1]{inputenc} 
\usepackage{graphicx}
\usepackage{amssymb}
\usepackage{cite}
\usepackage{amsbsy}
\usepackage{subfigure}
\usepackage{amsmath}
\usepackage{amsfonts}
\usepackage{multirow}
\usepackage{array}
\usepackage{lineno}
\usepackage{bm}
\usepackage[table]{xcolor}

\usepackage[normalem]{ulem}
\usepackage [latin1]{inputenc}
\usepackage{amsmath,amsfonts,amssymb}

\usepackage[colorlinks,linkcolor=black,anchorcolor=gray,citecolor=blue,urlcolor=blue]{hyperref}

\usepackage{cleveref}
\usepackage{algorithm}
\usepackage{booktabs}
\usepackage{multirow}
\usepackage{bigstrut}
\usepackage{longtable}
\usepackage{graphicx}
\usepackage{epstopdf}
\usepackage{enumerate}
\usepackage{url}
\usepackage{bbding}

\usepackage{algorithm}
\usepackage{algpseudocode}

\begin{document}
\title{FastHyMix: Fast and Parameter-free Hyperspectral Image Mixed Noise Removal}
%
%
%

\author{Lina~Zhuang,~
\IEEEmembership{Member,~IEEE,} and
Michael K. Ng,~
\IEEEmembership{Senior Member,~IEEE}   
\thanks{
This paper has  received final approval for publication as a Regular Paper in the IEEE Transactions on Neural Networks and Learning Systems.
}
\thanks{This work 
was supported    by    the National Natural Science Foundation of China under Grant 42001287.
The work of M. K. Ng was partially supported by the 
Hong Kong Research Grant Council General Research Fund ((HKRGC GRF)  under Grant
12300218, Grant 12300519, Grant 17201020 and Grant 17300021. (\textit{Corresponding author: Michael K. Ng.})
}
\thanks{L. Zhuang is with the Department of Mathematics, The University of
Hong Kong, Hong Kong, China, and also with Shenzhen Institute of Research
and Innovation, The University of Hong Kong, Shenzhen 518057, China
(e-mail: linazhuang@qq.com).}
\thanks{M.   K.   Ng   is   with   the   Department   of Mathematics,  The  University  of  Hong  Kong,  Hong  Kong, China (e-mail: mng@maths.hku.hk).}
}

\markboth{IEEE Transactions on Neural Networks and Learning Systems}%
{Shell \MakeLowercase{\textit{et al.}}: Bare Demo of IEEEtran.cls for Journals}


\maketitle

\begin{abstract}
The decrease of the  widths   of spectral bands in hyperspectral imaging
leads to the decrease in signal-to-noise ratio (SNR) of measurements.
 The decreased SNR reduces the reliability of measured features or information extracted from hyperspectral images (HSIs). 
Furthermore, the image degradations linked with various mechanisms also result in different types of noise, such as Gaussian noise, impulse noise, deadlines, and stripes.
This  paper introduces a \textbf{fast} and parameter-free \textbf{hy}perspectral image \textbf{mix}ed noise removal method (termed \textbf{FastHyMix}), which characterizes the complex distribution of mixed noise by using a Gaussian mixture model and exploits two main characteristics of hyperspectral data, namely low-rankness in the spectral domain and high correlation in the spatial domain. The Gaussian mixture model enables us to make a good estimation of Gaussian noise intensity and the locations of sparse noise.
The proposed method takes advantage of the low-rankness using subspace representation and the spatial correlation of HSIs by adding a powerful deep  image prior, which is extracted from a neural denoising network.
 An exhaustive array of experiments  and comparisons with state-of-the-art denoisers were carried out.
  The experimental results show  significant improvement in both synthetic and real datasets.
  A MATLAB demo of this work is available
at \url{https://github.com/LinaZhuang}  for the
sake of reproducibility.
\end{abstract}

\begin{IEEEkeywords}
Hyperspectral image denoising, Hyperspectral image restoration, Gaussian mixture model, low-rank representation, plug-and-play, sparse representation.
\end{IEEEkeywords}

%
\IEEEpeerreviewmaketitle

\section{Introduction}

Hyperspectral cameras  measure  the radiation arriving  at  the sensor with high spectral resolution over a
sufficiently broad spectral band such that the acquired spectrum can be used
to uniquely characterize and identify any given material \cite{overview}.
Hyperspectral imaging has been used in earth remote sensing tasks, such as  object classification \cite{8295275,8750899,9136736,9130919}, detection of landcover change, and  anomaly detection \cite{9404853,9288702},  and plays an important role
 in a wide array of applications, such as astronomy, agriculture,  and surveillance.
 However, due to the decrease of the  widths  of spectral bands, hyperspectral cameras receive fewer photons and tend to acquire images with lower signal-to-noise ratio (SNR). The decreased SNR reduces the reliability of measured features or information extracted from hyperspectral images (HSIs) \cite{overview}. 
Therefore, hyperspectral image denoising is a fundamental  preprocessing 
 before further applications. 

The image degradations linked with various mechanisms also result in different types of noise, such as Gaussian noise, Poissonian noise, impulse noise, deadlines/stripes, and cross-track illumination variation. In this paper, we focus on the discussion of additive and signal-independent noise (namely, Gaussian noise,  impulse noise, and deadlines/stripes)   and  attack  hyperspectral mixed noise composed of these additive noise.

Hyperspectral mixed noise is usually removed by  
 exploiting the distinct characteristics of HSIs and noise. 
Due to the very high spectral-spatial   correlation, hyperspectral data is low-rank and sparse on transform domain representation (such as data-adaptive subspace \cite{FastHyDe}, gradient domain \cite{chang2013simultaneous,chang2017hyper}, Fourier domain \cite{jiang2007hybrid}, wavelet domain \cite{rasti2013hyperspectral}, and Cosine domain). 
Gaussian noise in hyperspectral data is independent and densely distributed in the original image space and the above-mentioned transform domains. Consequently, 
a large amount of Gaussian noise can be removed from observations effectively either by using low-rank matrix/tensor factorization (in  NMoG \cite{NMoG},  double-factor  regularized  low-rank  tensor  factorization  (LRTF-DFR) method \cite{LRTF-DFR}, non-local meets global (NG-meet) method \cite{he2019non}, and L1HyMixDe \cite{L1HyMixDe}),  by minimizing the rank of the underlying clean HSI (in   low-rank  matrix   recovery (LRMR) method   \cite{LRMR},  nonconvex
  regularized low-rank and sparse matrix decomposition (NonRLRS) method  \cite{8760524}, and   DLR \cite{9374571}), or by using sparse representation of the underlying HSI ( in Kronecker-basis-representation (KBR) method  \cite{KBR}, a spectral-spatial adaptive hyperspectral
total variation (SSAHTV)  method \cite{SSAHTV}, the first order roughness penalty (FORP) method \cite{rasti2013hyperspectral}, and the sparse
  representation and low-rank constraint (`Spa+Lr') method \cite{zhao2014hyperspectral}.
Tensor approximation technique has been drawing increasing attention from HSI processing community, for example, a coupled sparse tensor factorization (CSTF)-based approach \cite{dian_8359412} is introduced for fusing hyperspectral and multispectral images, which has achieved outstanding performance.
Compared with using low-rank matrix approximation, low-rank tensor representation of an HSI can finely preserve its  intrinsic structure,  exploit data correlation in 3 dimensions simultaneously instead of 2 dimensions, and lead to better performance of mixed noise removal. For example,  \cite{zheng2019mixed}
introduces a HSI denoising method 3DLogTNN by modelling the underlying HSI as a tensor with low-fibered rank.
The non-convex low-rank tensor approximation (NonLRTA) method \cite{LIN2021126342} characterizes the clean HSI component by using the $\epsilon$-norm, which is a non-convex surrogate to Tucker rank. 
A tensor subspace
representation (TenSR) based HSI denoising method \cite{TenSR} takes advantage of the low-tubal rankness of the HSI tensor.

Impulse noise, deadlines, and stripe noise are usually sparse distributed in spatial and spectral domains, thus they are modelled as an additive and sparse component in observation models. The sparsity of impulse noise, deadlines, and stripe noise in cost functions can be promoted by minimizing its   $\ell_1$ norm in LRTF-DFR \cite{LRTF-DFR}, by constraining the upper bound of the cardinality of noise component in LRMR \cite{LRMR}, and by introducing a nonconvex regularizer named as normalized $\varepsilon$-penalty in NonRLRS \cite{8760524}.
 
The    structures of HSIs and  noise have been studied to develop a number of hyperspectral denoising methods, which address Gaussian noise well, but is not robust to mixed noise. There are several reasons: a) Both clean HSIs and stripes are low-rank \cite{chang2016remote}, thus only a low-rank regularization imposed on HSIs is not sufficient to separate them (see more experimental results of LRMR \cite{LRMR} and NonRLRS \cite{8760524} in Sections~\ref{sec:exp_sim} and \ref{sec:exp_real}). b) Noise in real HSIs exhibits very complex statistical distributions, calling for a powerful tool to characterize its structure. c) Noise  type  in each HSI varies according to the imaging conditions, such as atmospheric environment or illumination. Therefore, the type,  intensity, and distribution of noise varies in each HSI, calling for robust and data-adaptive denoising methods.
To address these hurdles, we propose
an adaptive Gaussian mixture model to fit the distribution of mixed noise so that we can find
 the noise structure  and estimate  the  noise intensity.

\subsection{Related work}

HSIs can be well approximated by low-dimensional subspace representations and  are characterized by a high level of self-similarity \cite{overview,gao2014subspace,AdeHyDe,8718504,Dian2019,8948303,dian2020recent,dian2019nonlocal,miao2021hyperspectral}.  
The idea of regularizing the subspace representation coefficients of HSIs  underlies   state-of-the-art   Gaussian-denoisers.  We refer to representative work: GLF \cite{GLF}, FastHyDe \cite{FastHyDe}, and NG-meet \cite{he2019non}.
This paper extends this strategy to  address mixed noise. The challenge of this extension lies in the  estimation  of the spectral subspace.

In the scenario of mixed noise, a spectral subspace is estimated iteratively and jointly with subspace coefficients of the HSI, for example, in LRTF-DFR \cite{LRTF-DFR} and     SNLRSF \cite{cao2019hyperspectral}. 
Joint estimation of the subspace and  the  corresponding coefficients of the HSI usually produce poor estimates of the subspace when HSI is affected by sever mixed noise. This paper introduces a  strategy to  estimate   the subspace and  the  corresponding coefficients of the HSI separately, leading to a non-iterative and more effective method.

Coefficients of subspace representation  of an HSI are termed  \textit{eigen-images}, which  have very high spatial correlation. Therefore, remaining noise in the eigen-images can be further alleviated by image filtering. For example, single-band image denoisers, BM3D \cite{BM3D}, WNNM  \cite{WNNM}, weighted TV  are used in HSI denoising methods, FastHyDe \cite{FastHyDe}, NG-meet \cite{he2019non}, and LRTF-DFR \cite{LRTF-DFR}, respectively. In this paper, eigen-images are regularized with  a more powerful deep  image prior, as neural networks have shown impressive  performance in recovering clean natural images from noisy observations \cite{DnCNN,FFDNet,CBDNet}. 
 Some neural networks have been conceived specially for addressing hyperspectral image noise.  We refer to some representative networks, 
such as a spatial-spectral gradient network (SSGN) \cite{SSGN},
a CNN-based HSI denoising method HSI-DeNet \cite{HSI-DeNet},   a
novel deep spatio-spectral Bayesian posterior (DSSBP) framework
\cite{DSSBP}, and a  3D dual-attention denoising network (3D-ADNet) \cite{3DADNet}.
 As the performance of deep learning based denoisers    highly depends on the quality and quantity of training data, a
challenge of deep-denoisers is the lack of real HSIs that
can be used as training data or how to   simulate pairs of clean-noisy
images close to real HSIs.
To sidestep the lack of training HSIs,  this work takes advantage of the  similarity between HSIs and RGB/grayscale images. Both kinds of images are natural images, sharing common image structures, thus it is reasonable that a denoising network,  well-trained using vast amounts of RGB images, also works well for HSIs.
  This paper incorporates  a well-known single-band deep denoiser, FFDNet \cite{FFDNet}, into a mixed noise removal framework derived using traditional ML technique. This fall in the line of research, called plug-and-play technique \cite{6737048,chan2016plug,dian2020regularizing} or regularization by denoising (RED) framework \cite{romano2017little}.

\subsection{Contributions}

 The work aims to recover an underlying clean HSI from observations corrupted by additive mixed noise (containing Gaussian noise, impulse noise, and deadlines/stripes) by characterizing the statistical distribution of mixed noise using a Gaussian mixture model, and exploiting the low-rankness in the spectral domain and high correlation in the spatial domain of HSIs.
  Contributions of  this work are summarized as follows: 
 \begin{itemize}
 \item A noise estimation method is proposed for mixed noise by exploiting high spectral correlation of HSIs. The mixed noise is partitioned into Gaussian noise and sparse noise using a Gaussian mixture model fitted to the mixed noise. The partition enables us to make a good estimation of Gaussian noise intensity per band and the locations of sparse noise.
 \item The proposed  mixed noise removal method is   user-friendly and parameter-free by allowing parameters adaptive to specific images. That is the parameters can be  set
adaptively  to noise statistics.
 \item An image prior extracted from a state-of-the-art neural denoising network, FFDNet, is seamlessly embedded within our HSI mixed noise removal framework, which is a successful combination of traditional machine learning technique and deep learning technique. Experimental results demonstrate that the embedded deep image prior significantly improve the estimation accuracy of clean HSIs.
 \end{itemize}

This paper   is organized as follows. Section \ref{sec:pro}
formulates the hyperspectral mixed noise removal problem. Section \ref{sec:noise} describes  a new noise estimation approach elaborated for mixed noise. Section \ref{sec:Fasthymix}
formally
introduces  the proposed mixed noise removal method. 
Sections \ref{sec:exp_sim} and \ref{sec:exp_real}  show and analyse the experimental results of the proposed method and the comparison methods. Finally, we make a conclusion of  this  paper in Section \ref{sec:concl}. 

\section{Problem formulation}
\label{sec:pro}

\begin{table*}[htbp]
  \centering
  \caption{Notations and Definitions}
    \begin{tabular}{cl}
    \toprule
    Notation & \multicolumn{1}{c}{Definition} \\
    \midrule
    ${\cal X} \in \mathbb{R}^{I_1 \times I_2   \times I_3}$ & $3$-dimensional tensor (calligraphic letter)\\
    ${\bf X}$    &  Matrix (boldface capital letter) \\
      ${\bf x}$     &  Vector (boldface lowercase letter) \\
      $x$     &  Scalar (italic lowercase letter)\\
    mode $n$ & $n$th dimension of a tensor\\
  mode-$3$ vectors of ${\cal X}$     
    & $I_3$-dimensional vectors obtained from   ${\cal X}$ by varying the $3$rd index   while   keeping the 1st and 2nd indices fixed.\\
   mode-3 slices of  $\cal{X}$ & Matrices obtained from $\cal{X}$ by fixing every index but the $3$rd index.  \\
  ${\cal X}(:, :, i)$    
     & $i$th mode-$3$ slice of ${\cal X}$, a  matrix obtained by   fixing the mode-$3$ index of ${\cal X}$ to be  $i$.\\
      ${\bf X}_{(3)} \in \mathbb{R}^{I_3 \times (I_1 *   I_2)}$ & Mode-$3$ unfolding of ${\cal X}$.  A tensor can be unfolded into a matrix by rearranging its mode-$3$ vectors, which are the column 
\\
&  vectors of ${\bf X}_{(3)}$.  \\
${\cal X}\times_3 {\bf E}$  & Tensor matrix multiplication. The mode-$3$ product of a tensor ${\cal X}  \in \mathbb{R}^{I_1 \times I_2   \times I_3}$ by a matrix ${\bf E} \in \mathbb{R}^{J_n \times I_3}$ is a tensor \\
&  ${\cal Y} \in \mathbb{R}^{I_1 \times    I_2 \times J_n}$, denoted as
${\cal Y} = {\cal X}\times_3 {\bf E}$,
which is  corresponding to a matrix multiplication, $ {\bf Y}_{(3)} = {\bf E} {\bf X}_{(3)}$.\\
$\| {\cal X} \|_F$ & The definition of Frobenius norm of a matrix is extended to a tensor as follows: $\| {\cal X} \|_F = \sqrt{\sum_{i_1, i_2, i_3}|x_{i_1, i_2, i_3}|^2}$ \\
    \bottomrule
    \end{tabular}%
  \label{tab:notations}%
\end{table*}%
 
Some notations and tensor operations  used in this paper and their definitions are provided in Tab. \ref{tab:notations}.
Let ${\mathcal X}  \in \mathbb{R}^{r \times c \times
B}$ denote  an underlying clean HSI with $r * c$   pixels and 
$B$ bands. Assuming that noise is additive, we can write  an observation model   as
\begin{equation}
\label{obs}
\mathcal{Y} = \mathcal{X} + \mathcal{G},
\end{equation}
where $\mathcal{Y},\mathcal{G}  \in \mathbb{R}^{r \times c \times B }$ denote an observed HSI data  and mixed  noise, respectively.
Elements  in  $\mathcal{G}$ are assumed to be a mixture of Gaussian noise, stripes, dealdlines, and impulse noise. Gaussian noise in real HSI tends to be non-independent and identically distributed
(non-i.i.d.) that is pixel-wise independent but band-wise dependent.

Due to the extreme high spectral correlation, hyperspectral vectors can be represented well in a low-dimensional subspace \cite{overview}, i.e., 
\begin{equation}
    \mathcal{X=Z} {\times_{3}}{\bf E},
\end{equation} 
with ${\bf E}\in\mathbb{R}^{B\times P}$ ($P\ll B$) and ${\cal Z} \in  \mathbb{R}^{r \times c \times
P} $.   $\bf E$ holds an orthogonal basis for the signal subspace, and  the  entries of ${\cal Z}$ are representation coefficients of ${\cal X}$ with respect to ${\bf E}$. Hereafter, mode-3 slices of ${\cal Z}$ are termed    the  \textit{eigen-images}.

\begin{figure*}[htbp]
\centering
\includegraphics[width=18cm]{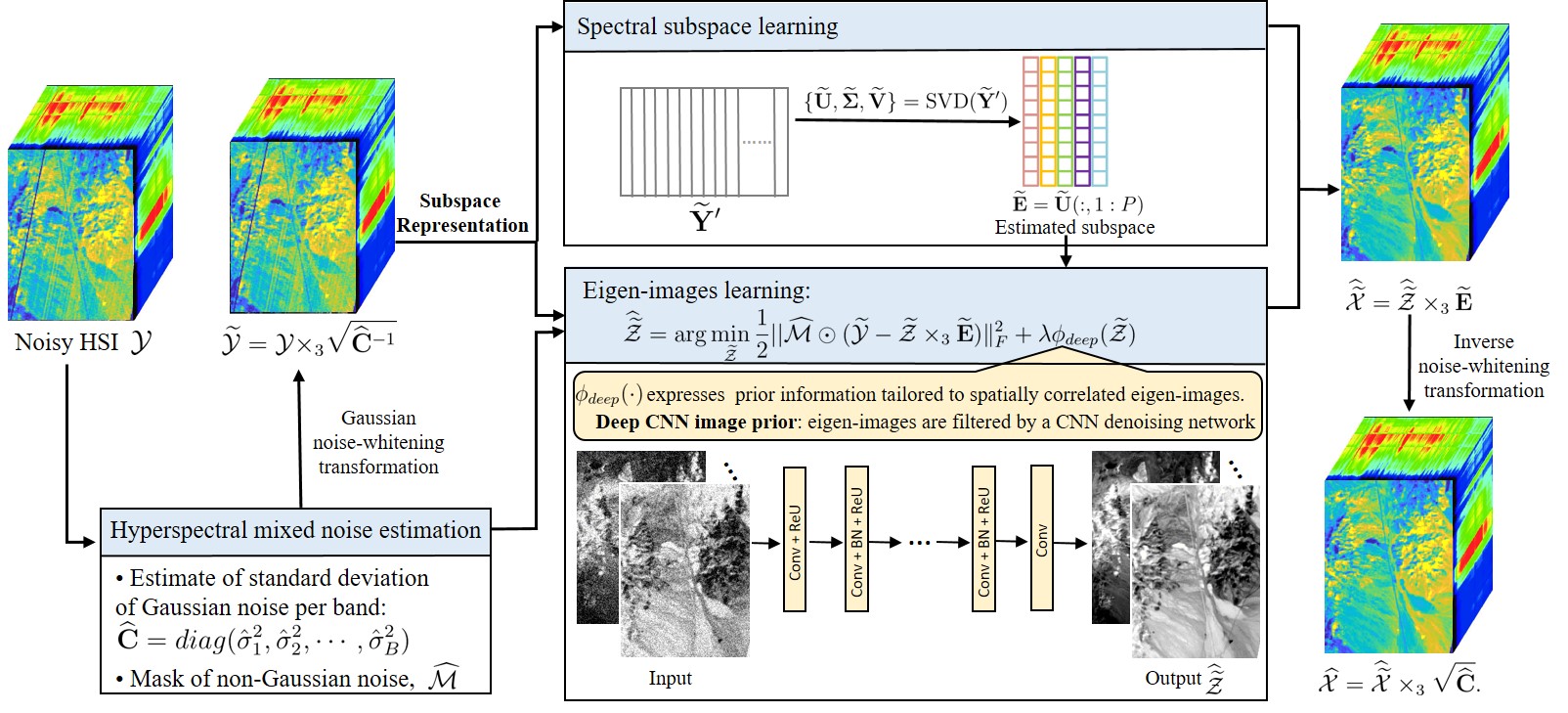}
\caption{Flowchart of the proposed mixed noise removal method, FastHyMix. }
\label{fig:flowchart_FastHyMix}
\end{figure*}

HSI mixed noise removal aims to estimate an underlying clean image $\mathcal{X}$,
given an observed noisy HSI $\mathcal{Y}$. 
This paper introduces  a \textbf{fast} and parameter-free \textbf{hy}perspectral image \textbf{mix}ed noise removal method (termed \textbf{FastHyMix}), which characterizes the complex distribution of mixed noise by using a Gaussian mixture model and exploits two main characteristics of hyperspectral data, namely low-rankness in the spectral domain and high correlation in the spatial domain. 
The main steps of the proposed method are described in the flowchart in Fig. \ref{fig:flowchart_FastHyMix}. 
Below we  first start by modelling 
the complex noise
as  a mixture of Gaussian densities, which is a universal
approximation to any continuous distribution and hence
capable of modelling a wide range of noise distributions \cite{8305626}. Then, we take advantage of the noise statistics learned from the fitted Gaussian mixture model, and derive a mixed noise removal algorithm adaptive to the  noise statistics of the image using subspace representation and deep image prior.

 \section{Hyperspectral mixed noise    estimation}
   \label{sec:noise}

\begin{figure*}[htbp]
\centering
\includegraphics[width=18cm]{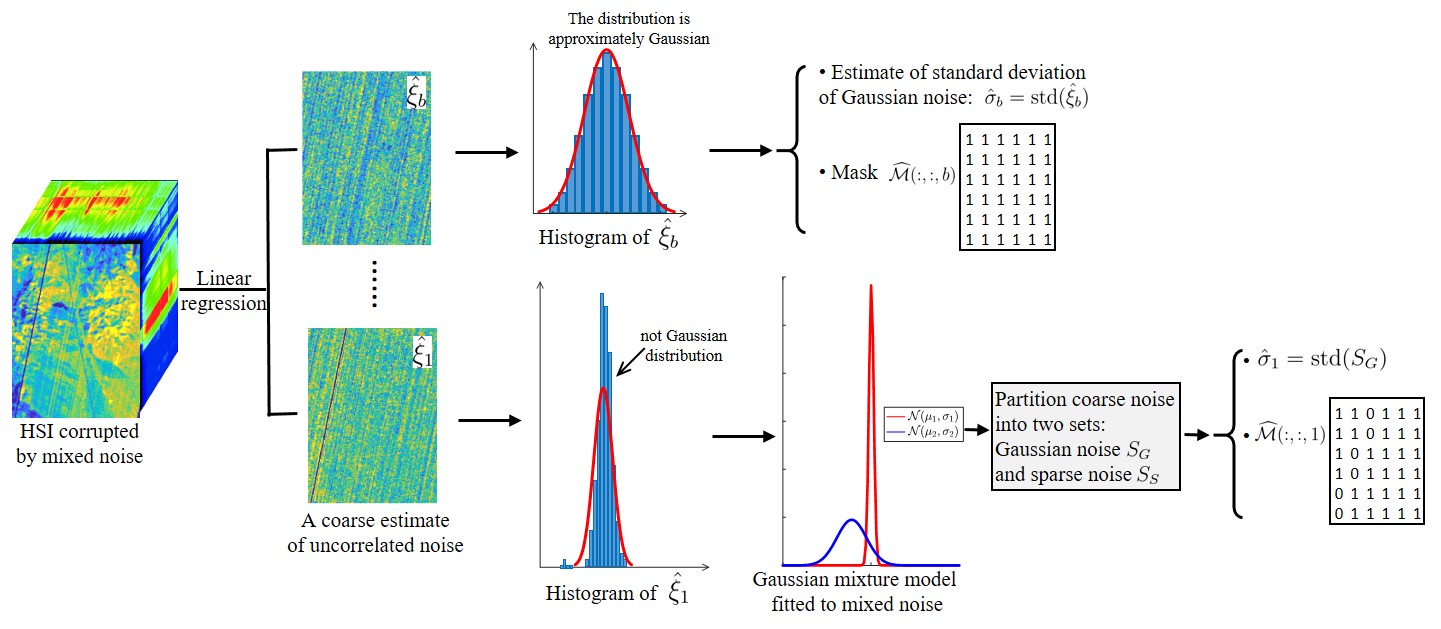}
\caption{ Flowchart of mixed noise estimation in an HSI.}
\label{fig:flowchart_noise_estimatioin}
\end{figure*}

In this section we  study the data structure of mixed noise by modelling the complex noise
as  a mixture of Gaussian densities, which enable us to make a good estimation of Gaussian noise intensity per band and the locations of sparse noise. 
Fig. \ref{fig:flowchart_noise_estimatioin} depicts the  procedure to  estimate noise statistics in HSIs. It performs band by band, and mainly contains following steps:
\begin{itemize}
\item[(a)] It starts by estimating a coarse noise per band by using a linear regression method exploiting spectral correlation of HSIs. 
\item[(b)] A normality test is used to determine if a histogram of coarse noise per band is well-modelled by a Gaussian distribution. 
\item[(c)] If a histogram of coarse noise  is approximately Gaussian, then we consider the noise in this band to be Gaussian noise.
\item[(d)] If a histogram of coarse noise  can not be fitted well by a  Gaussian distribution, then we consider the noise in this band to be a mixture of Gaussian noise and sparse noise. The mixed noise is then fitted by a Gaussian mixture model with 2 components. The fitting results   allow us to partition  the coarse noise  into two sets (namely, Gaussian noise and sparse noise). 
\item[(e)]  Estimate the standard deviation of Gaussian noise per band and the locations of sparse noise.
\end{itemize} 
Details of each step are given below.

  \subsection{Estimation of noise statistics}

Since spectral bands are highly correlated in a HSI, we assume that
 one band can be approximately represented as a linear combination of the remaining $(B -1)$ bands \cite{hysime,gao2013comparative}, that is
\begin{equation}
[{{\bf Y}}^T_{(3)}]_{:,b} = [{{\bf Y}}^T_{(3)}]_{\partial_{b}}\bm{\beta}_{b} + {\bm \xi}_{b},
\end{equation}
where ${{\bf Y}}^T_{(3)}$ denotes   the transpose of the mode-3 unfolding matrix ${{\bf Y}}_{(3)}$, the subscript $[\cdot]_{:,b}$ means extracting $b$th column from a matrix, a matrix with the subscript $[\cdot]_{\partial_{b}}$ means the matrix including all columns except $b$th column, ${\bm \beta}_{b} \in \mathbb{R}^{B-1}$ denotes regression coefficients, and   ${\bm \xi}_{b} \in \mathbb{R}^I$ (with $I = r * c$)  denotes regression error.

 The regression coefficients ${\bm \beta}_{b} $ can be estimated by the least squares method, i.e., 
\begin{equation}
\begin{array}{rl}
\hat{\bm \beta_{b}} &= \arg \underset{{\bm \beta}_{b}}{\mathop{  \min}}  \| [{{\bf Y}}^T_{(3)}]_{:,{b}} - [{{\bf Y}}^T_{(3)}]_{\partial_{b}}\bm{\beta}_{b}  \|_F^2 \\
& = 
([{{\bf Y}}^T_{(3)}]_{\partial_{b}}^T [{{\bf Y}}^T_{(3)}]_{\partial_{b}})^{-1}[{{\bf Y}}^T_{(3)}]_{\partial_{b}}^T [{{\bf Y}}^T_{(3)}]_{:,{b}}.  
\end{array}
\end{equation}
 Given $\hat{\bm \beta_{b}}$, the estimate of regression error, $\hat{\bm \xi}_{b}$,  is computed by
\begin{equation}
\hat{\bm \xi}_{b} = [{{\bf Y}}^T_{(3)}]_{:,{b}} - [{{\bf Y}}^T_{(3)}]_{\partial_{b}} \hat{\bm{\beta}}_{b}.
\end{equation}

We take the regression error, $\hat{\bm \xi}_{b}  (b=1, \cdots, B)$, as a coarse estimate of   uncorrelated noise in the $b$th band.
Statistical distribution of $\hat{\bm \xi}_{b}$ is studied band by band below.
 \subsection{Standard deviation estimation of Gaussian noise }
 
If the histogram of the coarse noise in ${b}$th band, $\hat{\bm \xi}_{b}$, can be approximated by a Gaussian distribution,  then we consider the coarse noise in this band to be Gaussian noise. Consequently, the standard deviation of Gaussian noise, denoted by $\hat{\sigma}_{b}$, is estimated by computing
\begin{equation}
\hat{\sigma}_{b} = \text{std}(\hat{\bm \xi}_{b}).
\end{equation}
where $\text{std}(\cdot)$ is a function computing  standard deviation of the elements of the input vector.
For normality test, we use  skewness and kurtosis estimates. 
That is if   absolute value of the skewness of $\hat{\bm \xi}_{b} $ is less than 3 and absolute value of its kurtosis  
is less than 10 \cite{bai2005tests}, then we consider $\hat{\bm \xi}_{b} $ to be approximately Gaussian distributed.
Otherwise, $\hat{\bm \xi}_{b} $ is a mixture of different kinds of noise.

\begin{figure}[htbp]
\centering
\includegraphics[width=8cm]{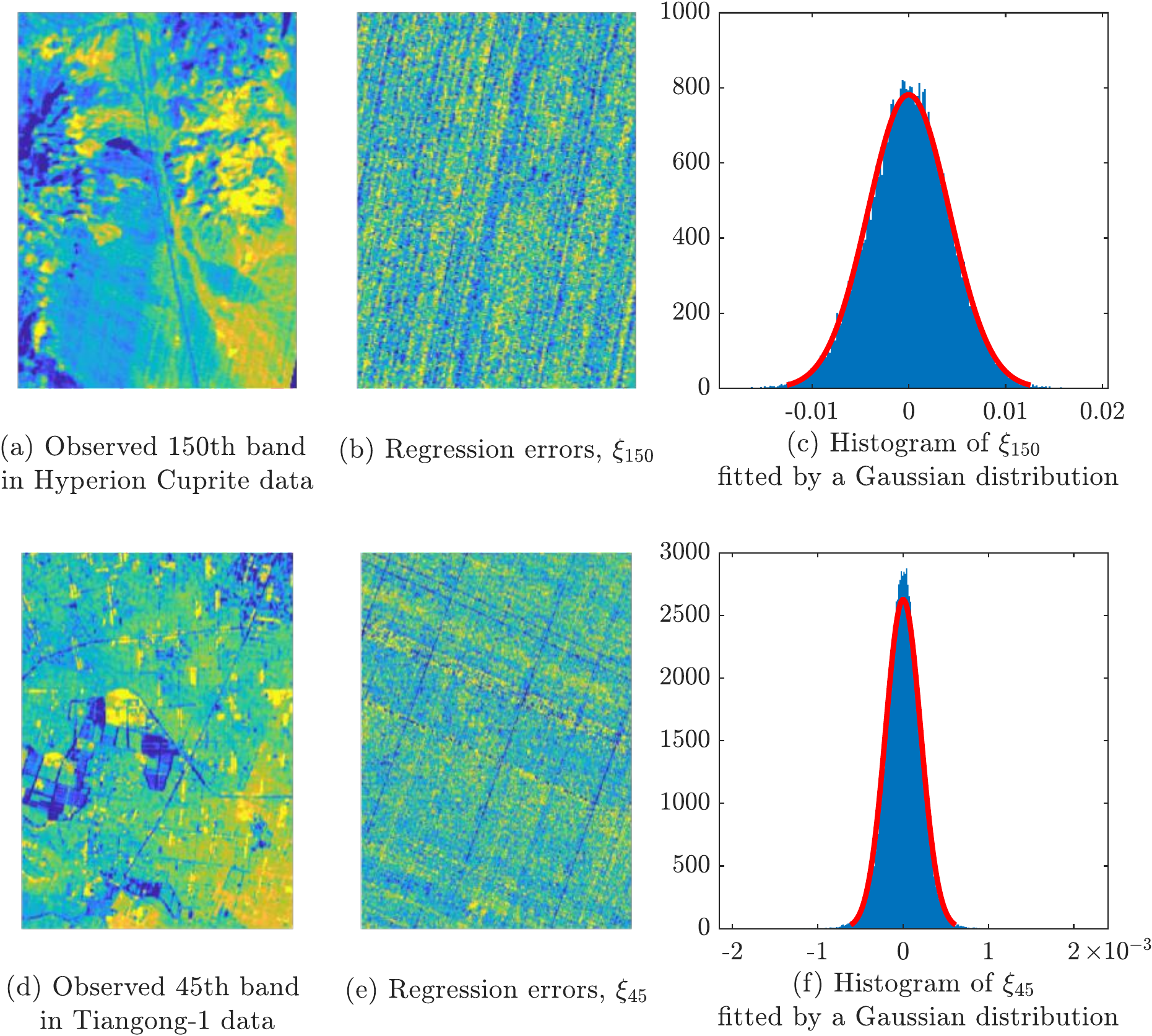}
\caption{The first row shows the 150th band in a real HSI: Hyperion Cuprite  image, and the second row shows the 45th band in a real HSI: Tiangong-1 image.}
\label{fig:Histogram_Gaussian}
\end{figure}

Two bands from real HSIs and their coarse noise are shown in Fig. \ref{fig:Histogram_Gaussian}, where we can see the histograms of coarse noise can be fitted well by Gaussian distributions. Therefore, we consider these two bands to have only Gaussian noise.

\begin{figure*}[htbp]
\centering
\includegraphics[width=18cm]{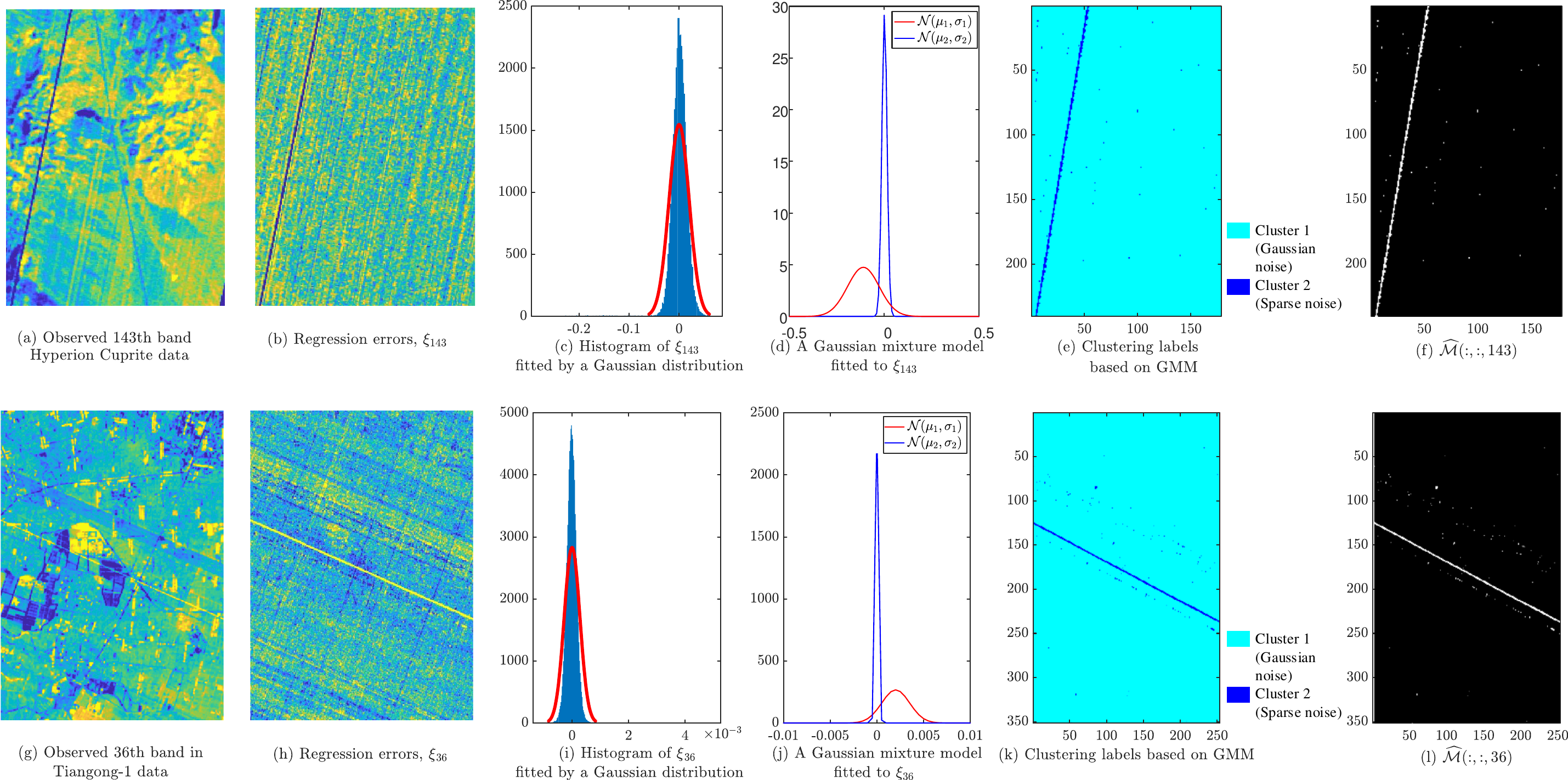}
\caption{The first row shows the 143th band of the Hyperion Cuprite  image and its noise statistics, and the second row shows the 36th band of the Tiangong-1 image and its noise statistics.}
\label{fig:Histogram_SparseNoise}
\end{figure*}

\subsection{Detection of pixels corrupted by mixed noise}

 \begin{figure}[htbp]
\centering
\includegraphics[width=7cm]{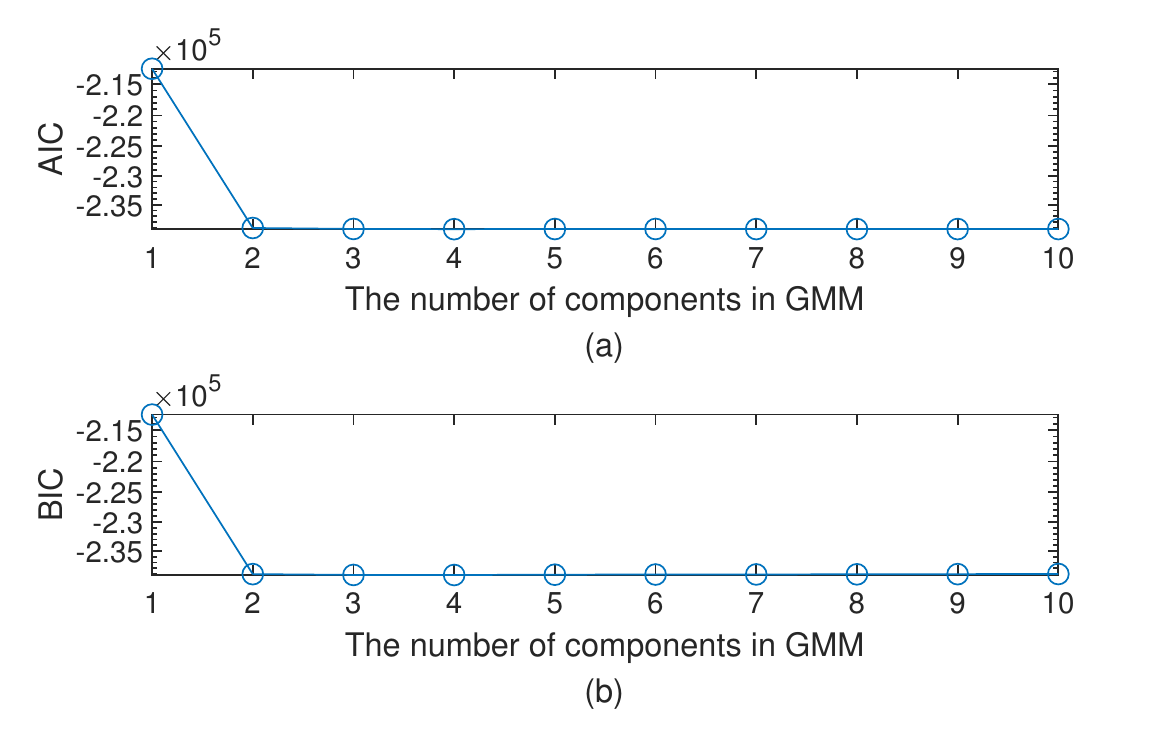}
\caption{AIC and BIC for various choices of the number of components in GMMs fitted to the coarse noise, $\xi_{143}$, in Hyperion Cuprite data.
}
 \label{fig:AIC_BIC}
\end{figure}

If the histogram of the coarse noise $\hat{\bm \xi}_{b}$ 
is not well-modelled by a Gaussian distribution, then it implies that the noise in the ${b}$th band is a mixture of Gaussian noise and sparse noise. 
For example, Fig. \ref{fig:Histogram_SparseNoise} shows two bands from two real HSIs and their corresponding coarse noise. The red curves in Fig. \ref{fig:Histogram_SparseNoise}-(c,i)  are  Gaussian distributions fitted to the histograms of coarse noise. Obviously, simple Gaussian distributions do not fit the coarse noise well and underfitting occurs.
 Then,  we fit this mixed noise using a Gaussian mixture model, which is a universal approximation to any continuous distribution and hence
capable of modelling a complex noise distribution. 
  To create a useful GMM, the choose of the number of components (i.e., the number of Gaussian distributions) should be careful.
We can choose the best number of components using Akaike Information Criterion (AIC) \cite{AIC} and Bayesian Information Criterion (BIC) \cite{BIC}. 
The AIC or BIC for a model is usually written in the form $(-2\log  L + cK)$, where $L$ is the likelihood function, $K$ is the number of parameters in the model (i.e., the number of components in GMMs), and $c$ is 2 for AIC and $\log(I)$ for BIC.
Both AIC and BIC are a way to find the balance between a good fit and over complexity in a model.
Fig. \ref{fig:AIC_BIC} shows the AIC and BIC scores of GMMs with various number of components fitted to the coarse noise shown in Fig. \ref{fig:Histogram_SparseNoise}-(b). 
The model with 2 components is the best one that balances low AIC and BIC with simplicity.
Also, AIC and BIC have been tested in other images  in the experimental part and it comes to a conclusion that the GMMs with 2 components are good enough to fit the data.
Therefore, the number of components in GMM is set to 2   in our work.
 Meanwhile, for HSI mixed noise removal problem, we mainly interest in the statistical parameters of Gaussian noise and the spatial locations of sparse noise. 
A GMM with 2 components can exactly group  elements in $\hat{\bm \xi}_{b}$    into two sets: Gaussian noise and sparse noise.

 Let ${\bf a}_b = [ a_{1b}, \cdots, a_{ib}, \cdots, a_{Ib}]^T := {\bm \xi}_{b} \in \mathbb{R}^I$ denotes the coarse noise in ${b}$th band. 
 To simplify the notation in the following derivation, we omit its band index  in the next subsection. Then we have ${\bf a} = [ a_{1}, \cdots, a_{i}, \cdots, a_{I}]^T := {\bm \xi}_{b} $.

\subsubsection{Gaussian mixture model fitted to mixed noise}

A Gaussian mixture distribution model   with 2 components fitted to the mixed noise takes the form
\begin{equation}
f(a_i, {\bf \Psi}) = \sum_{k=1}^2 \pi_k {\cal N}(a_i; \mu_k, \sigma_k^2),
\end{equation}
 where $\pi_k$ denotes nonnegative and sum-to-one mixing proportions, ${\cal N}(a_i; \mu_k, \sigma_k^2)$
 are the component Gaussian densities with mean value $\mu_k$ and variance $\sigma_k^2$ computed at $a_i$, and ${\bf \Psi} = (\mu_1,  \mu_2, \sigma_1^2, \sigma_2^2, \pi_1, \pi_2)^T$ denotes a vector of unknown parameters.

The maximum likelihood estimate of ${\bf \Psi}$ is written as
\begin{equation}
\begin{array}{rl}
\widehat{\bf \Psi} &=  \arg \underset{{\bf \Psi}}{\mathop{ \max}} \log  L({\bf \Psi} ) \\
&  = \arg \underset{{\bf \Psi}}{\mathop{ \max}} \sum_{i=1}^{I} \log\{ \sum_{k=1}^2 \pi_i {\cal N}(a_i; \mu_k, \sigma_k^2) \}.
\end{array} 
\end{equation}
 The above   maximum likelihood problem can be solved by implementing Expectation-maximization (EM) algorithm \cite{EM}, which starts by conceptualizing $a_1, \cdots, a_{I}$ to have arisen from one of the component distributions of the mixture model. We define an indicator, $ u_{ik} =1$ if $a_i$ has arisen  from the $k$th component distribution, and $ u_{ik} = 0$ otherwise.
Let ${\bf \Psi}^{(t)} := \{ \mu_1^{(t)}, \mu_2^{(t)}, \sigma_1^{(t)}, \sigma_2^{(t)}, \pi_1^{(t)}, \pi_2^{(t)} \}$ denote the parameters of mixture distributions on the $t$th iteration of the EM algorithm.
The EM algorithm seeks to find the maximum likelihood estimate  of mixture distributions by iteratively applying the following two steps (see \cite{EM} for details).

a) E-step: Given ${\bf \Psi}^{(t)}$, E-step on the $(t+1)$th iteration calculates the current conditional expectation of $U_{ik}$ given data $a_i$, where $U_{ik}$ is the random variable corresponding to $u_{ik}$. We  have
\begin{equation}
{\mathbb E}(U_{ik} | a_i, {\bf \Psi}^{(t)}) = \tau_k (a_i; {\bf \Psi}^{(t)}),
\end{equation}
where
\begin{equation}
 \tau_k (a_i; {\bf \Psi}^{(t)})  = \frac{\pi_k^{(t)} {\cal N}(a_i | \mu_k^{(t)}, \sigma_k^{2(t)})}{f(a_i, {\bf{\Psi}}^{(t)})}.
\end{equation}
After taking the conditional expectation with ${\bf \Psi} = {\bf \Psi}^{(t)}$, we have 
\begin{equation}
\begin{array}{rl}
&Q( {\bf \Psi}; {\bf \Psi}^{(t)} ) = \\
&\sum_{k=1}^2 \sum_{i=1}^{I} \tau_k(a_i; {\bf \Psi}^{(t)}) \{ \log \pi_k + \log {\cal N}(a_i | \mu_k, \sigma_k^{2}) \}.
\end{array}
\end{equation}

b) M step: The M-step on the $(t+1)$th iteration updates ${\bf \Psi}$ by calculating the global maximization of $Q({\bf \Psi}; {\bf \Psi}^{(t)})$ with respect to $\pi_k,~\mu_k,~\sigma_k~(k=1,2)$, yielding
\begin{equation}
\pi_k^{(t+1)} = \sum_{i=1}^{I} \tau_k (a_i; {\bf \Psi}^{(t)})/n,~~~~(k=1,~2),
\end{equation}
\begin{equation}
\mu_k^{(t+1)} = \frac{\sum_{i=1}^{I} \tau_k(a_i; {\bf \Psi}^{(t)}) a_i }{ \sum_{i=1}^{I} \tau_k(a_i; {\bf \Psi}^{(t)})},~~~~(k=1,~2),
\end{equation}
and
\begin{equation}
\sigma_k^{2(t+1)} = \frac{ \sum_{i=1}^{I} \tau_k(a_i; {\bf \Psi}^{(t)}) (a_i - \mu_k^{(t+1)})^2  }{\sum_{i=1}^{I} \tau_k(a_i; {\bf \Psi}^{(t)}) },~~~~(k=1,~2).
\end{equation}

\subsubsection{Cluster using Gaussian mixture model}

Given $\widehat{\bf \Psi}$, the posterior probability that the element $a_i$ arose from group $k$ is given by the fitted posterior probability,
\begin{equation}
\tau_k(a_i; \widehat{\bf \Psi}) = \frac{ \hat{\pi}_k {\cal N}(a_i; \hat{\mu}_k, \hat{\sigma}_k^2)}{ f(a_i, \widehat{{\bf \Psi}}) }~~(k=1,2; i=1, \cdots, I).
\end{equation} 

The coarse noise elements $(a_1, \cdots, a_I)$ can be partitioned into two sets ${\bf S}=\{ S_{G}, S_{S} \}$ (where $S_{G}$ and $S_{S}$ represent a set of Gaussian noise and a set of sparse noise, respectively)  by assigning each $a_i$ to the group to which it has the highest estimated posterior probability of belonging. Let $L_i \in \{1, 2\}$ denotes the cluster label of $a_i$. We have 
\begin{equation}
L_i   = \{k | k= \arg \max_k \tau_k (a_i; \widehat{\bf \Psi}) \}.
\end{equation}

As Gaussian noise is densely distributed and other kinds of noise (namely, stripes, deadlines, and impulse noise) is usually sparse distributed in the spatial domain, 
the group with larger component proportion is considered to be Gaussian noise, and the other group is sparse noise (see Fig. \ref{fig:Histogram_SparseNoise}-(e,k)). 

\subsection{Estimation of ${\bf C}$ and ${\cal M}$} %
After clustering the coarse noise using the Gaussian mixture model, we can derive  the following noise statistics, which are of importance for conceiving a parameter-free denoising algorithm. 
\begin{itemize}
\item The standard deviation of Gaussian noise in the ${b}$th band can be computed using the set of Gaussian noise   via $\hat{\sigma}_{b} = \text{std}(S_{Gb})$, where $S_{Gb}$ represents a set of coarse noise elements in the $b$th band belonging to sparse noise. Then, an estimate of the covariance matrix of Gaussian noise is obtained as $\widehat{\bf C} = diag( \hat{\sigma}_1^2, \cdots, \hat{\sigma}_{b}^2, \cdots, \hat{\sigma}_{B}^2 )\in \mathbb{R}^{{B} \times{B}}$. 
\item  Let $\mathcal{M}\in {{\{0,\text{ }1\}}^{r\times c \times {B}}}$ denotes  a mask tensor indicating noise types as follows:

${{m}_{i,j,b}}=\left\{ \begin{array}{ll}
   1,& \text{when observation }{{y}_{i,j,b}}\text{ is corrupted only } \\
   & \text{by Gaussian noise;}\\
   0, & \text{when }{{y}_{i,j,b}}\text{ is corrupted by mixed noise.} \\
\end{array}\right. $

We can derive an estimate of the mask, $\widehat{\cal M}$, from the clustering result of each element in the coarse noise (see Fig. \ref{fig:Histogram_SparseNoise}-(f,l)).
\end{itemize}

   \section{FastHyMix: Fast HSI mixed noise removal}
   \label{sec:Fasthymix}
Given the estimates of the correlation matrix of Gaussian noise and the locations of sparse noise, we introduce a fast HSI mixed noise removal method, preceded by a noise-whitening step, which is based on the estimated correlation matrix of Gaussian noise, $\widehat{\bf C}$.

  \subsection{Gaussian noise-whitening transformation} 

Gaussian noise in real HSIs usually has different intensity per band. To remove Gaussian noise
effects on spectral subspace learning, we include a noise-whitening process. Let $\widetilde{\cal Y} \in \mathbb{R}^{r \times c \times B}$ denote the noise-whitened image,  and we have
  \begin{equation}
\widetilde{\cal Y} =  {\cal Y}  {\times_3} \sqrt{\widehat{\bf C}^{-1}} =\widetilde{\cal X}  + \widetilde{\cal O},
\end{equation}
where $\widetilde{\cal X} =  {\cal X}  {\times_3} \sqrt{\widehat{\bf C}^{-1}}$ and  $\widetilde{\cal O} = {\cal O} {\times_3} \sqrt{\widehat{\bf C}^{-1}}$.
After whitening, mode-3 vectors of $\widetilde{\cal X}$ still live in a low-dimensional subspace, thus
  \begin{equation}
\widetilde{\cal Y} = \widetilde{\cal Z} {\times_{3}}\widetilde{\bf E}  + \widetilde{\cal O},
\end{equation}
where columns of $\widetilde{\bf E} \in \mathbb{R}^{B \times P}$ span an orthogonal subspace, which can be estimated from the observations corrupted only by Gaussian noise. 
We take observed pixels corrupted only by Gaussian noise from $\widetilde{{\bf Y}}$ and stack them as columns to make a single matrix, denoted as
  $\widetilde{{\bf Y}}' \in \mathbb{R}^{B \times Q}$ with $Q$ pixels. An estimate of the spectral subspace can be obtained as
\begin{equation}
\label{eq:learnE}
\widetilde{\bf E} = \widetilde{\bf U}(:,1:P),
\end{equation}
where $P$ denotes the dimension of spectral subspace,   $\widetilde{{\bf U}} \in \mathbb{R}^{B \times B}$ is an orthogonal matrix and  $\{ \widetilde{\bf U}, \widetilde{\bf \Sigma}, \widetilde{\bf V}  \} = \text{SVD}(\widetilde{{\bf Y}}')$ with singular values in $\widetilde{\bf \Sigma}$ ordered  by  non-increasing  magnitude. 
In the scenario of mixed noise, the estimation of dimension of signal subspace is challenging \cite{overview,hysime}. Fortunately, our proposed method is   extremely robust
to errors in estimation of the subspace dimension, as far as the subspace
dimension is not underestimated. Relative evidence and analysis are provided in Section~\ref{sec:robustness}.

\subsection{Estimation of eigen-images}
Given the estimate of the spectral subspace, the subspace coefficients of the HSI can be estimated by solving the following optimization problem:
   \begin{equation}
\label{eq:opt}   
\widehat{\widetilde{{\mathcal{Z}}}}=\arg \underset{\widetilde{\mathcal{Z}}}{\mathop{ \min }}\,\frac{1}{2}||\widehat{\mathcal{M}}\odot (\widetilde{\mathcal{Y}}-\widetilde{\mathcal{ Z}} \times_{3} {\widetilde{\bf E}})\|_{F}^{2}+\lambda \phi_{deep} (\widetilde{\mathcal{Z}}),
\end{equation}
where $\odot$ denotes element-wise multiplication, and $\lambda >0$ is a parameter of the regularization.
The first term on the right hand side represents the data fidelity and accounts only for the Gaussian noise, and $\phi_{deep}(\cdot)$  is a  regularizer expressing prior information tailored to spatially correlated eigen-images.

Solvers for optimization problem (\ref{eq:opt}) will be iterative due to a non-diagonal operator involved in (\ref{eq:opt}). To sidestep iterations and speed up the algorithm, 
we  propose a suboptimal
solution that is very fast and effective.
Problem (\ref{eq:opt}) 
is solved approximately by the following two steps.

In the  first step, we estimate the  components corrupted by sparse noise at each pixel, $\tilde{{\bf y}}_i\in \mathbb{R}^{B}~(i=1, \cdots, I)$,  by solving a simple least square problem:
\begin{equation}
\label{eq:learn_z}
\begin{array}{rl}
{\bar{\tilde{\bf z}}}_i& = \arg \min_{\tilde{\bf z}_i}  \| \hat{\bf m}_i \odot( \tilde{{\bf y}}_i - \widetilde{\bf E}\tilde{\bf z}_i) \|_2\\
  & = (\widetilde{\bf E}^T \text{diag}(\hat{\bf m}_i) \widetilde{\bf E})^{-1} \widetilde{\bf E}^T(\hat{\bf m}_i  \odot \tilde{\bf y}_i),~~(i=1, \cdots, I),
\end{array}
\end{equation}
where  $\tilde{{\bf y}}_i \in \mathbb{R}^{B}$,
 $\hat{\bf m}_i \in \mathbb{R}^{B}$,  $\tilde{\bf z} \in \mathbb{R}^{P}$ are $i$th   pixel in the noise-whiten image,  corresponding mask, and subspace coefficients, respectively. 
The matrix $(\widetilde{\bf E}^T \text{diag}(\hat{\bf m}_i) \widetilde{\bf E})$ is non-singular when $\|\hat{\bf m}_i\|_0 \ge P$, meaning   the number of  components corrupted by only Gaussian noise
is larger or equal than the dimension of the signal subspace.
 Given ${\bar{\tilde{\bf z}}}_i$, we can recover the components corrupted by sparse noise 
   at each pixel by computing
\begin{equation}
\label{eq:learn_y}
{\bar{\tilde{\bf y}}}_i = \widetilde{\bf E} {\bar{\tilde{\bf z}}}_i,~~(i=1, \cdots, I).
\end{equation}

Now remaining noise in ${\bar{\tilde{\bf y}}}_i~(i=1, \cdots, I)$ is mainly Gaussian,
thus
in the second step, we perform a Gaussian noise removal step on $\bar{\widetilde{\cal Y}} := \text{folding}( [ {\bar{\tilde{\bf y}}}_1, \cdots, {\bar{\tilde{\bf y}}}_{B}])\in \mathbb{R}^{r \times c \times B}$: 
\begin{align}
\label{eq:solveZ2} 
\widehat{\widetilde{\cal Z}} &= \arg  \underset{{\widetilde{\cal Z}}} {\mathop{ \min}} \frac{1}{2} \| {\bar{\widetilde{\cal Y}}} -  \widetilde{\cal Z}\times_3 \widetilde{\bf E}  \|_F + \lambda \phi_{deep}(\widetilde{\cal Z})\\
\label{eq:solveZ} 
 & = \arg  \underset{{\widetilde{\cal Z}}} {\mathop{ \min}} \frac{1}{2} \| {\bar{\widetilde{\cal Y}}} \times_3 \widetilde{\bf E}^T -  \widetilde{\cal Z}  \|_F + \lambda \phi_{deep}(\widetilde{\cal Z}),
  \end{align}
which is a proximity operator of $\phi_{deep}$ applied to $\widetilde{\cal Z}' :=  {\bar{\widetilde{\cal Y}}} \times_3 \widetilde{\bf E}^T$. 
A proof of the equivalence of \eqref{eq:solveZ2} and \eqref{eq:solveZ} can be found in Appendix A of \cite{GLF}. 
Considering that the orthogonal projection is a decorrelation transformation and mode-3 slices of   of $\widetilde{\cal Z}$  tend to be decorrelated, we decouple $\phi_{deep}(\cdot)$ with respect  to the mode-3 slices,
that is 
\begin{equation}
  \phi_{deep}(\widetilde{\cal Z}) = \sum_{p=1}^{P} \phi_{deep,p} ( \widetilde{\cal Z}(:,:,p)),
\end{equation}
where $\widetilde{\cal Z}(:,:,p)$ denotes  the $i$th mode-3 slice.
The solution of \eqref{eq:solveZ} is decoupled w.r.t. $\widetilde{\cal Z}(:,:,p)$ and may be written as
\begin{equation}
\label{eq:Z_i}
\begin{array}{r}
\widehat{\widetilde{\cal Z}}(:,:,p)  = \arg  \underset{{\widetilde{\cal Z}}(:,:,p)} {\mathop{ \min}} \frac{1}{2} \|  \widetilde{\cal Z}'(:,:,p) -  \widetilde{\cal Z}(:,:,p)   \|_F \\
+ \lambda \phi_{deep,p}(\widetilde{\cal Z}(:,:,p)), ~p = 1, \dots, P,
\end{array}
\end{equation}
To solve subproblem \eqref{eq:Z_i}, we resort to the plug-and-play trick \cite{6737048,RhyDe,Fu2021,hy-demosaicing}, whose main idea is to
 directly use an existing regularizer from a state-of-the-art denoiser, instead of investing effort in designing a new regularizer exploiting the high spatial  correlation of eigen-images. In this paper, the prior of a  denoising network, FFDnet\footnote{\url{https://github.com/cszn/FFDNet}} \cite{FFDNet}, is plugged in \eqref{eq:Z_i}, leading to
\begin{equation}
\label{eq:ffdnet}
\widehat{\widetilde{\cal Z}}(:,:,p) \leftarrow \text{FFDNet}( \widetilde{\cal Z}'(:,:,p), \lambda ),
 \end{equation}
where the function FFDNet$(\cdot)$ outputs a denoised eigen-image.
In the FFDNet network, parameter $\lambda$ is related to standard deviation of Gaussian noise in the input image. As Gaussian noise in $\widetilde{\cal Y}$ has been whitened, we set $\lambda = 1$ in \eqref{eq:ffdnet}.
 We remark that other state-of-the-art single-band Gaussian-denoisers (such as BM3D \cite{BM3D} and WNNM  \cite{WNNM}) also can be adopted to estimate the eigen-images. The FFDNet is selected in this work due to its  following advantages over others. Compared with other machine learning-based denoisers \cite{BM3D,WNNM}, deep-learning-based FFDNet is much faster as long as it has been well trained. Compared with other deep-learning-based denoisers \cite{jain2008natural,HSI-DeNet}, FFDNet is able to address images with various noise levels, meaning that we can input a new image without retraining.

The clean data of $\widetilde{\cal Y}$ is recovered as
\begin{equation}
 \widehat{\widetilde{\cal X}} =  \widehat{\widetilde{\cal Z}} \times_3 \widetilde{\bf E}.
\end{equation}
\subsection{Inverse noise-whitening transformation}
Finally, we perform inverse noise-whitening transformation to obtain an estimate of the clean HSI:
\begin{equation}
\widehat{\cal X} =  \widehat{\widetilde{\cal X}} \times_3 \sqrt{\widehat{\bf C}}.
\end{equation}
   
The pseudocode in   Algorithm \ref{pseudocode} shows how FastHyMix  is implemented to reduce mixed noise for an HSI. 
 Given an HSI of  size $r$ (rows) $\times c$ (columns) $\times  B$ (bands) with subspace dimension $P$ ($P \ll B$), the computational complexity of obtaining $\widehat{\bf C}$ and $\widehat{\mathcal{M}}$ in line 1 is $\mathcal{O}(r*c*B^3)$ and $\mathcal{O}(r*c*B)$, respectively. The Gaussian noise-whitening   in line 2, its inverse transformation in line 6, and the image reconstruction step in line 5 have same computational complexity, that is  $\mathcal{O}(r*c*B^2)$. The estimation of the  spectral subspace in line 3 and eigen-images in line 4 respectively cost $\mathcal{O}(r^2*c^2*B)$ and $\mathcal{O}(r*c*B^2*P + P*d)$,  where $d$ represents the computational complexity of denoising an eigenimage.
 Consequently, the overall computational complexity of FastHyMix
is $\mathcal{O}(r*c*B^3 + r^2*c^2*B + P*d)$.

\begin{algorithm}[h]
\caption{FastHyMix: Fast Hyperspectral Image Mixed Noise Removal}
\begin{algorithmic}[1]
\renewcommand{\algorithmicrequire}{\textbf{Input:}} 
\renewcommand{\algorithmicensure}{\textbf{Output:}}
\Require
A noisy HSIb $\mathcal{Y}  $
\State Estimation of noise statistics: $\widehat{\bf C}$ and $\widehat{\cal M}$ 
\State Gaussian noise-whitening: $\widetilde{\cal Y} =  {\cal Y}  {\times_3} \sqrt{\widehat{\bf C}^{-1}}$ 
\State Estimation of the spectral subspace, $\widehat{\bf E}$, via \eqref{eq:learnE}
\State Estimation of eigen-images, $\widehat{\widetilde{\cal Z}}$, via \eqref{eq:learn_z}, \eqref{eq:learn_y}, and \eqref{eq:ffdnet}
\State $ \widehat{\widetilde{\cal X}} =  \widehat{\widetilde{\cal Z}} \times_3 \widetilde{\bf E}.$
\State Inverse noise-whitening: $\widehat{\cal X} =  \widehat{\widetilde{\cal X}} \times_3 \sqrt{\widehat{\bf C}}.$
\Ensure The denoised HSI $\widehat{\mathcal{X}}$
\end{algorithmic}
\label{pseudocode}
\end{algorithm}

\section{Experiments with simulated images}
\label{sec:exp_sim}
\begin{figure}[htbp]
\centering
\subfigure[Subregion of Washington DC Mall data]{
\includegraphics[width=0.11\textwidth]{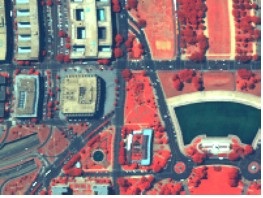}}
\subfigure[Subregion of Pavia University data]{
\includegraphics[width=0.11\textwidth]{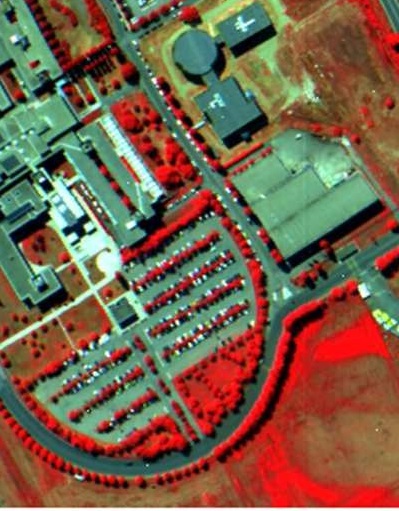}}
\subfigure[Hyperion Cuprite]{
\includegraphics[width=0.11\textwidth]{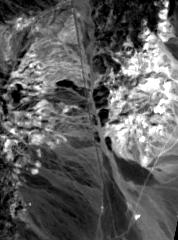}}
\subfigure[Tiangong-1]{
\includegraphics[width=0.11\textwidth]{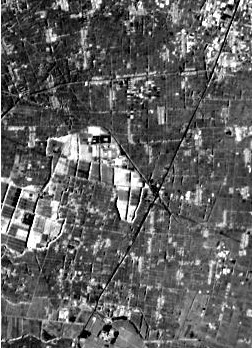}}
\caption{HSIs used in the experiments.}
\label{fig:RealImg}
\end{figure}

The experiments of hyperspectral mixed noise removal were conducted on two simulated hyperspectral datasets and two real hyperspectral datasets (see Fig. \ref{fig:RealImg}).

\subsection{Simulation of noisy datasets and comparisons}

Two noisy HSIs were generated based on two public hyperspectral
datasets (shown in Fig. \ref{fig:RealImg}-(a,b)), namely
a subregion of Washington DC Mall data\footnote{\url{https://engineering.purdue.edu/~biehl/MultiSpec/hyperspectral.html}}
(of size 150(rows) $\times$ 200(columns) $\times$ 191(bands))
and 
a subregion of Pavia University data\footnote{\url{http://www.ehu.eus/ccwintco/index.php?title=Hyperspectral_Remote_Sensing_Scenes}}
 (of size 310(rows) $\times$ 250(columns) $\times$ 87(bands)).
Firstly, we removed bands severely corrupted by water vapor in the atmosphere. 
To obtain relatively clean images, we projected spectral vectors of each image onto a subspace spanned by principle eigenvectors of each image. The projection of each image is considered to be  clean image.

To simulate noisy HSIs, we added four kinds of additive noise into images as follows:

\textbf{Case 1 (Gaussian non-i.i.d. noise)}: ${\bf n}_i \sim \mathcal{N}({\bf 0}, {\bf D}^2) $ where ${\bf D}$ is a diagonal matrix with diagonal elements sampled from a uniform distribution $U(0.01, 0.02)$ in Washington DC Mall data and a uniform distribution $U(0.05, 0.10)$ in Pavia University data.

 \textbf{ Case 2 (Gaussian noise + stripes)}: Synthetic data with Gaussian noise (described in case 1) and oblique stripe noise randomly affecting 30\% of the bands and, for each band, about random 10\%  of the pixels.

\textbf{ Case 3 (Gaussian noise + `Salt \& Pepper' noise)}: Synthetic data with Gaussian noise (described in case 1) and `Salt \& Pepper' noise with noise density 0.5\%, meaning affecting approximately 0.5\% of elements in ${\cal X}$.

\textbf{ Case 4 (Gaussian noise + stripes + `Salt \& Pepper' noise)}: Synthetic data with Gaussian noise (described in case 1), random oblique stripes (described in case 2) and `Salt \& Pepper' noise (described in case 3).

Six recent state-of-the-art hyperspectral image denoising methods are taken for experimental comparison. They are  NG-meet \cite{he2019non}\footnote{\url{https://github.com/quanmingyao/NGmeet}},  LRMR  \cite{LRMR}\footnote{\url{https://sites.google.com/site/rshewei/home}}, NonRLRS \cite{8760524}, KBR \cite{KBR}\footnote{\url{http://gr.xjtu.edu.cn/web/dymeng/}}, NMoG \cite{NMoG}, and   LRTF-DFR  method \cite{LRTF-DFR}\footnote{\url{https://yubangzheng.github.io/homepage/\#publications}}.  Among them, the NG-meet is conceived to address hyperspectral Gaussian noise, but we included it to see whether mixed noise could be removed well by a Gaussian-denoiser. 
 All experiments were implemented in MATLAB (R2020a) on Windows 10 with an 
ADM Ryzen 9 4900HS 3.00-GHz 
 processor and 16-GB RAM.
 
Regarding the parameter setting of compared methods, we basically   fine-tuned the parameters of regularizations for all simulated and real images.
Also, for fair comparison, we set same values for   common parameters, such as  the dimension of spectral subspace used in NG-meet, LRTF-DFR, and FastHyMix.

For quantitative assessment, the peak signal-to-noise ratio (PSNR) index,  the structural similarity (SSIM) index,   the feature similarity
(FSIM) index of each band, and the spectral angle distance (SAD) were calculated.
The SSIM metric \cite{SSIM}, measuring the similarity between two images, is  considered  to  be  correlated  with  
the  quality  perception  of  the  human  visual  system  (HVS). The SSIM  is  designed  by  modeling  any  image  distortion  as  a  
combination  of  three  factors  that  are  loss  of  correlation,  luminance  distortion,  and  contrast  distortion.
The FSIM metric \cite{FSIM} measures
the similarity of images using gradient magnitude and Fourier phase congruency.
 The mean PSNR (MPSNR), mean SSIM (MSSIM),   mean FSIM (MFSIM),  and mean SAD (MSAD) over bands of denoised images are presented in Tab.~\ref{tab:twoSimulation}, where we highlighted the best results in bold.

\begin{table*}[htbp]
  \centering
  \caption{Performance of the Proposed and Comparison Methods on  Washington DC Mall Data and Pavia University Data}
    \begin{tabular}{cccccccccc}
    \toprule
          & \multicolumn{1}{c}{Indexes} & \multicolumn{1}{c}{Noisy} & \multicolumn{1}{c}{NG-meet} & \multicolumn{1}{c}{LRMR} & \multicolumn{1}{c}{NonRLRS} & \multicolumn{1}{c}{KBR} & \multicolumn{1}{c}{NMoG} & \multicolumn{1}{c}{LRTF-DFR} & \multicolumn{1}{c}{FastHyMix} \\
          &       &       & \multicolumn{1}{c}{\cite{he2019non} } & \multicolumn{1}{c}{\cite{LRMR}} & \multicolumn{1}{c}{\cite{8760524} } & \multicolumn{1}{c}{\cite{KBR}} & \multicolumn{1}{c}{\cite{NMoG}} & \multicolumn{1}{c}{\cite{LRTF-DFR}} & \multicolumn{1}{c}{(Proposed)} \\
    \midrule
    \multicolumn{10}{c}{Simulated Washington DC Mall data} \\
    \midrule
    \multirow{5}[2]{*}{Case 1} & MPSNR (dB) & 19.33 & 37.48 & 34.13 & 31.74 & 33.84 & 32.36 & 39.73 & \textbf{40.20 } \\
          & MSSIM & 0.8586 & 0.9973 & 0.9941 & 0.9884 & 0.995 & 0.9907 & 0.9981 & \textbf{0.9982 } \\
          & MFSIM & 0.9017 & 0.9937 & 0.989 & 0.9833 & 0.9905 & 0.9892 & \textbf{0.998} & \textbf{0.9980 } \\
          & MSAD & 0.124  & 0.016  & 0.022  & 0.027  & 0.024  & 0.036  & \textbf{0.014 } & \textbf{0.014 } \\
          & Time (s) & 0     & 28    & 21    & 19    & 115   & 53    & 65    & \textbf{2 } \\
    \midrule
    \multirow{5}[2]{*}{Case 2} & MPSNR (dB) & 12.62 & 12.79 & 26.42 & 24.63 & 31.24 & 32.34 & 35.58 & \textbf{37.90 } \\
          & MSSIM & 0.7589 & 0.6356 & 0.974 & 0.9643 & 0.9894 & 0.9907 & 0.9922 & \textbf{0.9970 } \\
          & MFSIM & 0.8351 & 0.8462 & 0.9733 & 0.9592 & 0.9847 & 0.9892 & 0.9939 & \textbf{0.9959 } \\
          & MSAD & 0.584 & 0.348  & 0.071  & 0.132  & 0.030 & 0.036  & 0.033  & \textbf{0.018 } \\
          & Time (s) & 0     & 30    & 19    & 39    & 117   & 53    & 81    & \textbf{6 } \\
    \midrule
    \multirow{5}[2]{*}{Case 3} & MPSNR (dB) & 13.61 & 31.08 & 34.12 & 32.07 & 31.25 & 32.19 & 35.64 & \textbf{38.97 } \\
          & MSSIM & 0.8248 & 0.9885 & 0.9941 & 0.989 & 0.9899 & 0.9906 & 0.9936 & \textbf{0.9978 } \\
          & MFSIM & 0.8877 & 0.9833 & 0.9891 & 0.9838 & 0.9847 & 0.9892 & 0.9934 & \textbf{0.9975 } \\
          & MSAD & 0.177  & 0.028  & 0.022  & 0.026  & 0.030  & 0.036 & 0.024  & \textbf{0.015 } \\
          & Time (s) & 0     & 31    & 20    & 25    & 114   & 59    & 68    & \textbf{7 } \\
    \midrule
    \multirow{5}[2]{*}{Case 4} & MPSNR (dB) & 8.46  & 13.59 & 26.51 & 24.58 & 33.02 & 32.1  & 35.53 & \textbf{37.22 } \\
          & MSSIM & 0.7291 & 0.682 & 0.976 & 0.9642 & 0.9936 & 0.9906 & 0.9922 & \textbf{0.9966 } \\
          & MFSIM & 0.8236 & 0.8501 & 0.9729 & 0.9593 & 0.9897 & 0.9891 & 0.9938 & \textbf{0.9949 } \\
          & MSAD & 0.602  & 0.326  & 0.070  & 0.130  & 0.025  & 0.036  & 0.033  & \textbf{0.018 } \\
          & Time (s) & 0     & 30    & 20    & 40    & 118   & 52    & 78    & \textbf{7 } \\
    \midrule
    \multicolumn{10}{c}{Simulated Pavia University data} \\
    \midrule
    \multirow{5}[2]{*}{Case 1} & MPSNR (dB) & 20.52 & 36.5  & 34.27 & 30.87 & 33.5  & 35.43 & 32.88 & \textbf{38.83 } \\
          & MSSIM & 0.3499 & 0.9328 & 0.9181 & 0.9068 & 0.9162 & 0.9339 & 0.8475 & \textbf{0.9749 } \\
          & MFSIM & 0.6671 & 0.9671 & 0.9639 & 0.9491 & 0.9605 & 0.9737 & 0.933 & \textbf{0.9845 } \\
          & MSAD & 0.540  & 0.124  & 0.103  & 0.114  & 0.111  & 0.087  & 0.189  & \textbf{0.055 } \\
          & Time (s) & 0     & 70    & 30    & 22    & 124   & 76    & 104   & \textbf{3 } \\
    \midrule
    \multirow{5}[2]{*}{Case 2} & MPSNR (dB) & 18.28 & 28.4  & 32.73 & 30.15 & 33.36 & 35.41 & 32.71 & \textbf{38.54 } \\
          & MSSIM & 0.3259 & 0.7799 & 0.9169 & 0.8964 & 0.9164 & 0.9317 & 0.8443 & \textbf{0.9742 } \\
          & MFSIM & 0.6386 & 0.8926 & 0.9571 & 0.9412 & 0.9599 & 0.9732 & 0.9324 & \textbf{0.9840 } \\
          & MSAD & 0.668  & 0.329  & 0.116  & 0.146  & 0.111  & 0.088  & 0.189  & \textbf{0.056 } \\
          & Time (s) & 0     & 72    & 27    & 32    & 129   & 73    & 129   & \textbf{6 } \\
    \midrule
    \multirow{5}[2]{*}{Case 3} & MPSNR (dB) & 18.61 & 33.3  & 34.23 & 30.67 & 33.48 & 34.24 & 32.86 & \textbf{38.63 } \\
          & MSSIM & 0.3124 & 0.866 & 0.9218 & 0.9017 & 0.9157 & 0.9275 & 0.8461 & \textbf{0.9746 } \\
          & MFSIM & 0.6432 & 0.9399 & 0.9644 & 0.9468 & 0.9602 & 0.9681 & 0.9323 & \textbf{0.9843 } \\
          & MSAD & 0.602 & 0.187  & 0.099  & 0.120  & 0.111  & 0.099  & 0.192  & \textbf{0.055 } \\
          & Time (s) & 0     & 73    & 30    & 22    & 128   & 78    & 126   & \textbf{12 } \\
    \midrule
    \multirow{5}[2]{*}{Case 4} & MPSNR (dB) & 16.79 & 28.21 & 32.78 & 29.9  & 33.33 & 35.33 & 32.75 & \textbf{38.28 } \\
          & MSSIM & 0.2911 & 0.7689 & 0.9165 & 0.8921 & 0.9152 & 0.9313 & 0.8456 & \textbf{0.9736 } \\
          & MFSIM & 0.6175 & 0.8875 & 0.9571 & 0.9388 & 0.9597 & 0.973 & 0.9328 & \textbf{0.9836 } \\
          & MSAD& 0.709  & 0.336  & 0.115  & 0.145  & 0.112  & 0.088  & 0.192  & \textbf{0.057 } \\
          & Time (s) & 0     & 72    & 30    & 34    & 129   & 78    & 108   & \textbf{12 } \\
    \bottomrule
    \end{tabular}%
   \label{tab:twoSimulation}%
\end{table*}%

 \subsection{Mixed noise removal}
 
 \begin{figure*}[htbp]
\centering
\includegraphics[width=17cm]{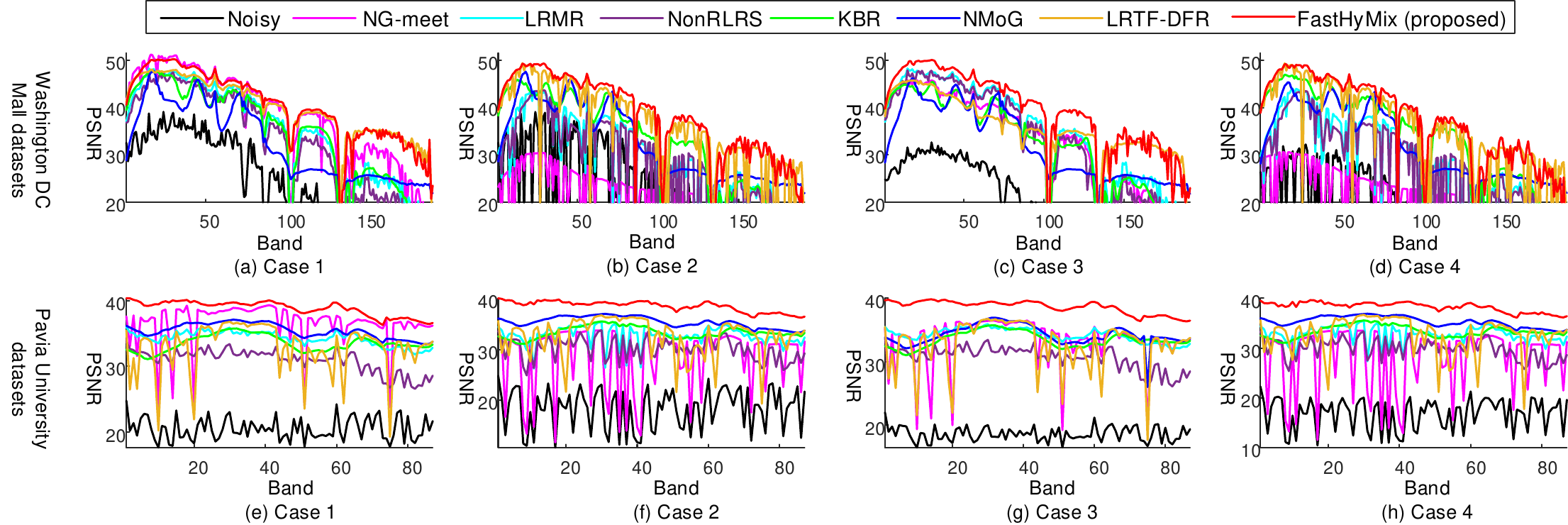}
\caption{Band-wise PSNR values of  denoised Washington DC Mall data  in the first row and of denoised  Pavia University data  in the second row. Subfigures in (a,e), (b,f), (c,g), (d,h) correspond to case 1, case 2, case 3, and case 4, respectively.}
\label{fig:psnr_curves}
\end{figure*}

 \begin{figure*}[htbp]
\centering
\includegraphics[width=16cm]{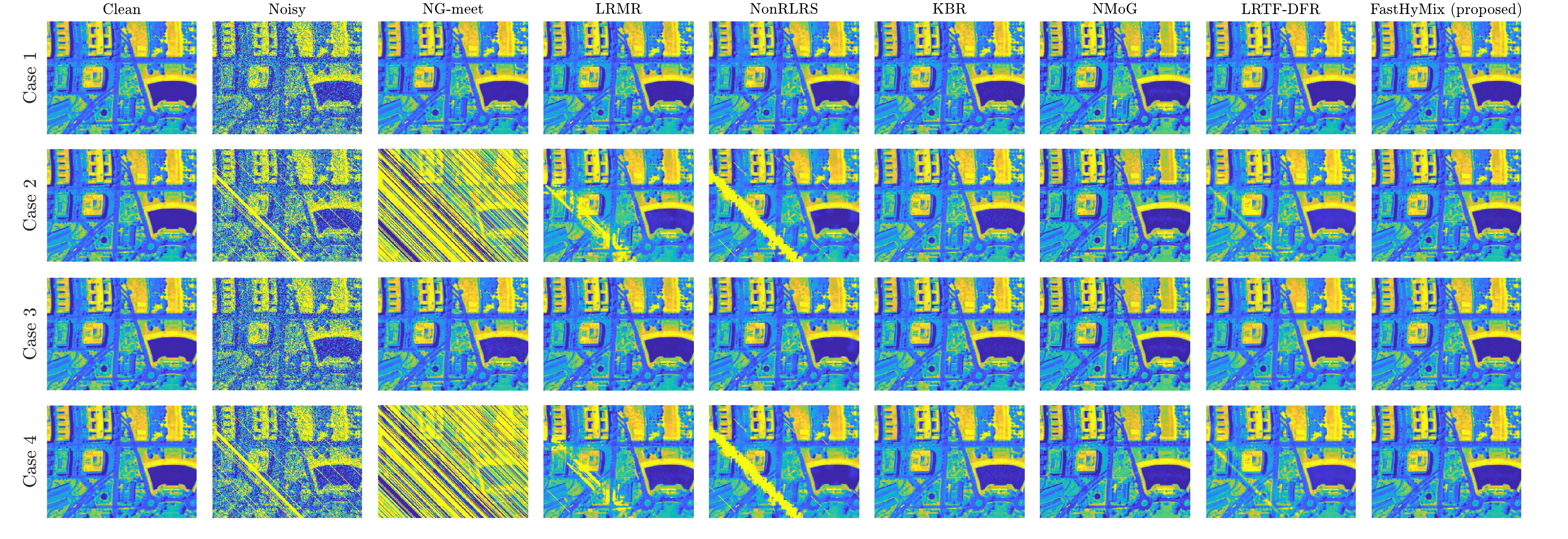}
\caption{Band 126 of the simulated Washington DC Mall data before and after denoising in four cases.}
\label{fig:bands_DC_126}
\end{figure*}

\begin{figure*}[htbp]
\centering
\includegraphics[width=16cm]{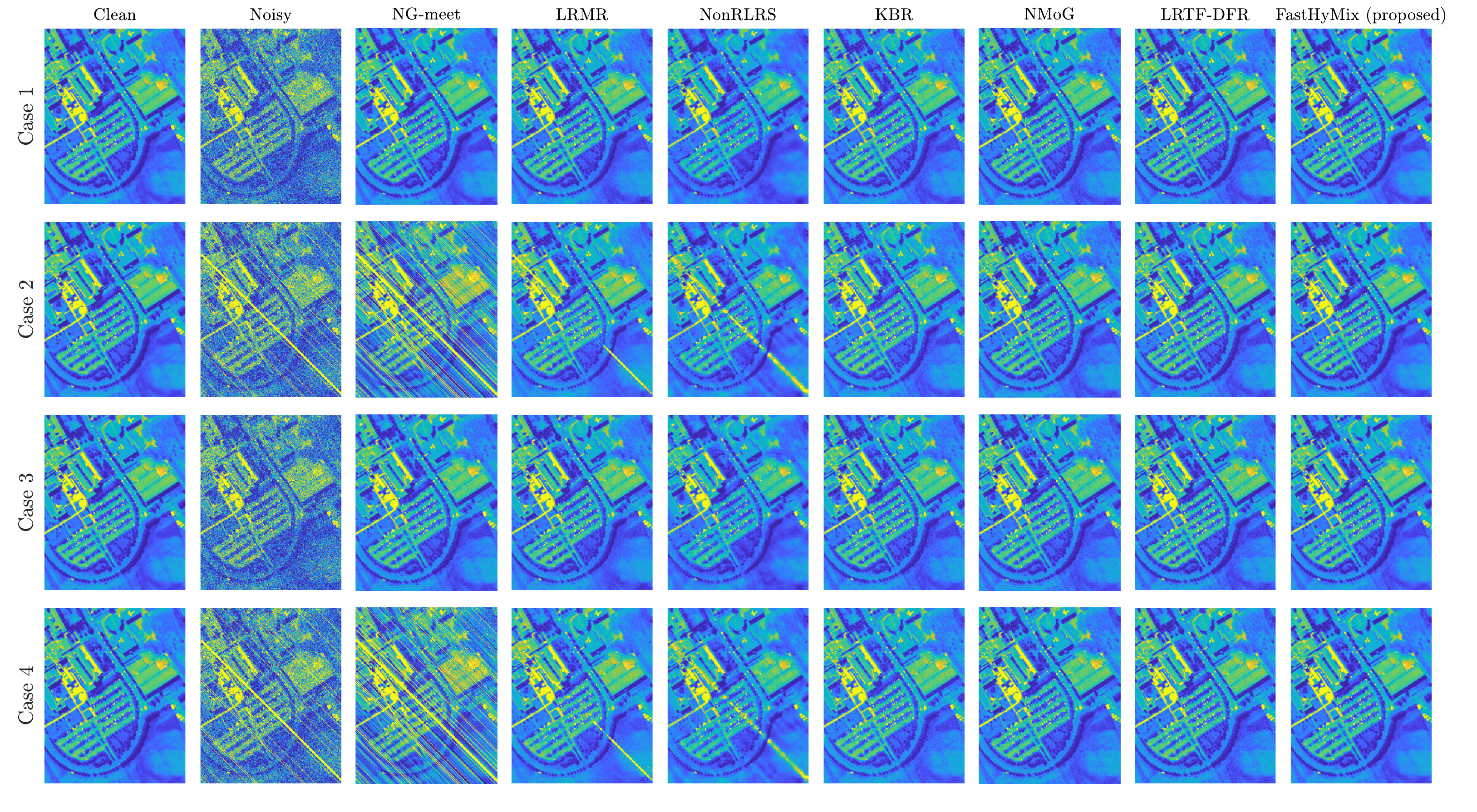}
\caption{Band 37 of the simulated Pavia University data before and after denoising in four cases.}
\label{fig:bands_pavia_37}
\end{figure*}

 Mixed noise removal performance of each method on two simulated datasets  in terms of MPSNR, MSSIM,   MFSIM, and MSAD is presented in Tab. \ref{tab:twoSimulation}. 
  The band-wise PSNR  is depicted in Fig. \ref{fig:psnr_curves}  for quantitative assessment. 
   It can be seen that FastHyMix uniformly yeilds the best performance  in the shortest time in HSIs with different kinds of noise.  Among the competitors, NG-meet is conceived specially for addressing Gaussian noise, thus it works well in case 1 (including only Gaussian noise), but not in cases 2-4 (including mixed noise). The results of NG-meet in four cases imply that a mixture of noise cannot be addressed simply using a Gaussian-denoiser, thus call for efficient mixed noise removal methods. 
 High correlation between spectral channels of HSIs leads to low-rank  structure of the HSIs in the spectral domain, which is exploited  by constraining the upper bound of the rank in LRMR, by introducing a nonconvex normalized $\varepsilon$-penalty to the matrix rank in NonRLRS, by relaxing the tensor rank term with a log-sum form in KBR, and by low-rank matrix/tensor factorization in NMoG, LRTF-DFR, and FastHyMix. Results in Tab. \ref{tab:twoSimulation} show that low-rank matrix/tensor factorization-based denoisers achieve better denoising performance.

For visual comparison, we display 126th band of Washington DC Mall data and
37th band of Pavia University data      in Figs. \ref{fig:bands_DC_126} and  \ref{fig:bands_pavia_37}, respectively.  For case 1 (Gaussian noise), all methods can reduce noise significantly. 
 As shown in   Figs. \ref{fig:bands_DC_126} and  \ref{fig:bands_pavia_37},     LRMR,    NonRLRS, and  LRTF-DFR,   are able to remove light stripes, but still leaving  some wide stripes. Heavy stripe noise still remains in the results of NG-meet. But we emphasize that it is unfair to compare with NG-meet in the cases including mixed noise, as NG-meet is designed specially for Gaussian noise. 
KBR, NMoG,  and FastHyMix methods visually yield comparable results in  Figs. \ref{fig:bands_DC_126} and  \ref{fig:bands_pavia_37}.

\subsection{  Estimation of mixed noise}

 \begin{figure}[htbp]
\centering
\includegraphics[width=8cm]{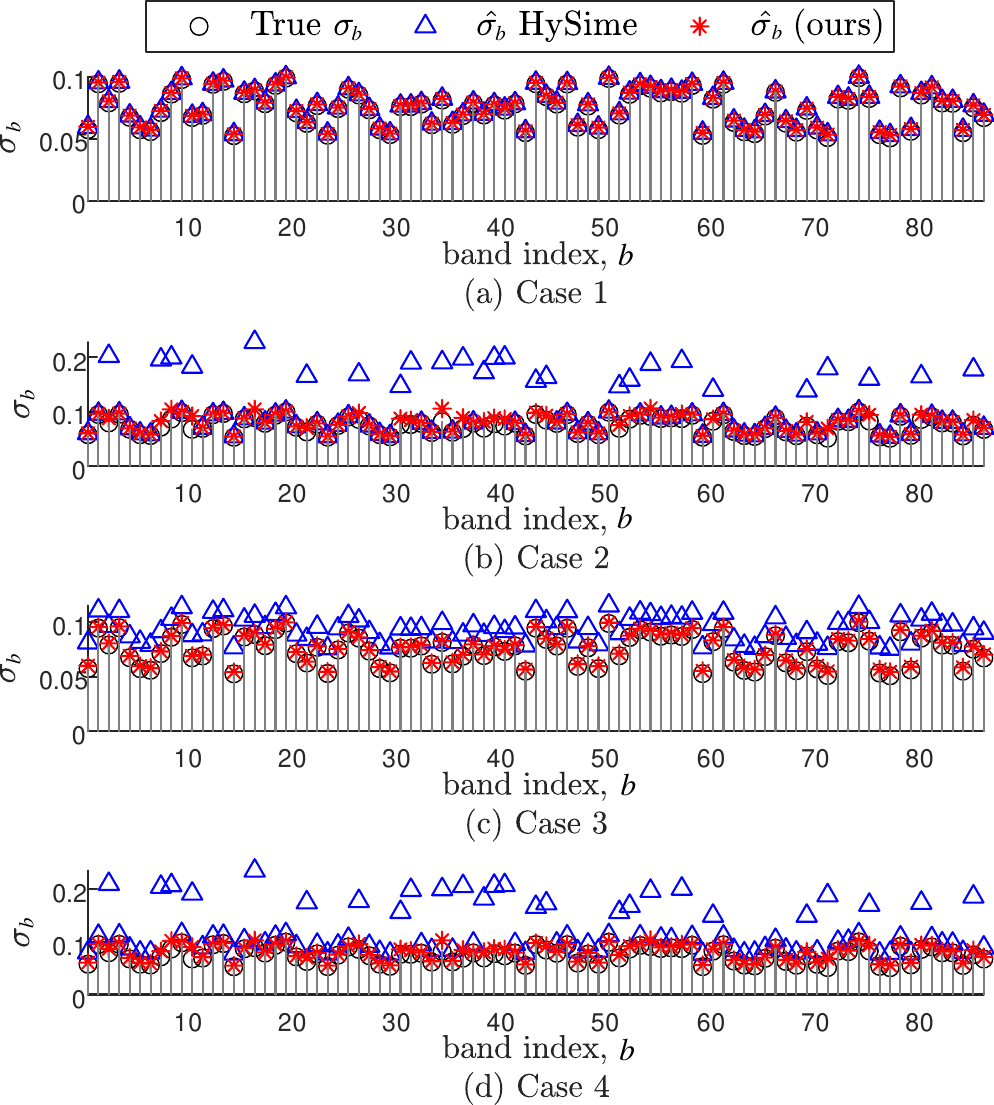}
\caption{Estimates of standard deviations of Gaussian noise per band in simulated Pavia University dataset.  }
\label{fig:std_Gaussian_comparison}
\end{figure}

\begin{figure}[htbp]
\centering
\includegraphics[width=6cm]{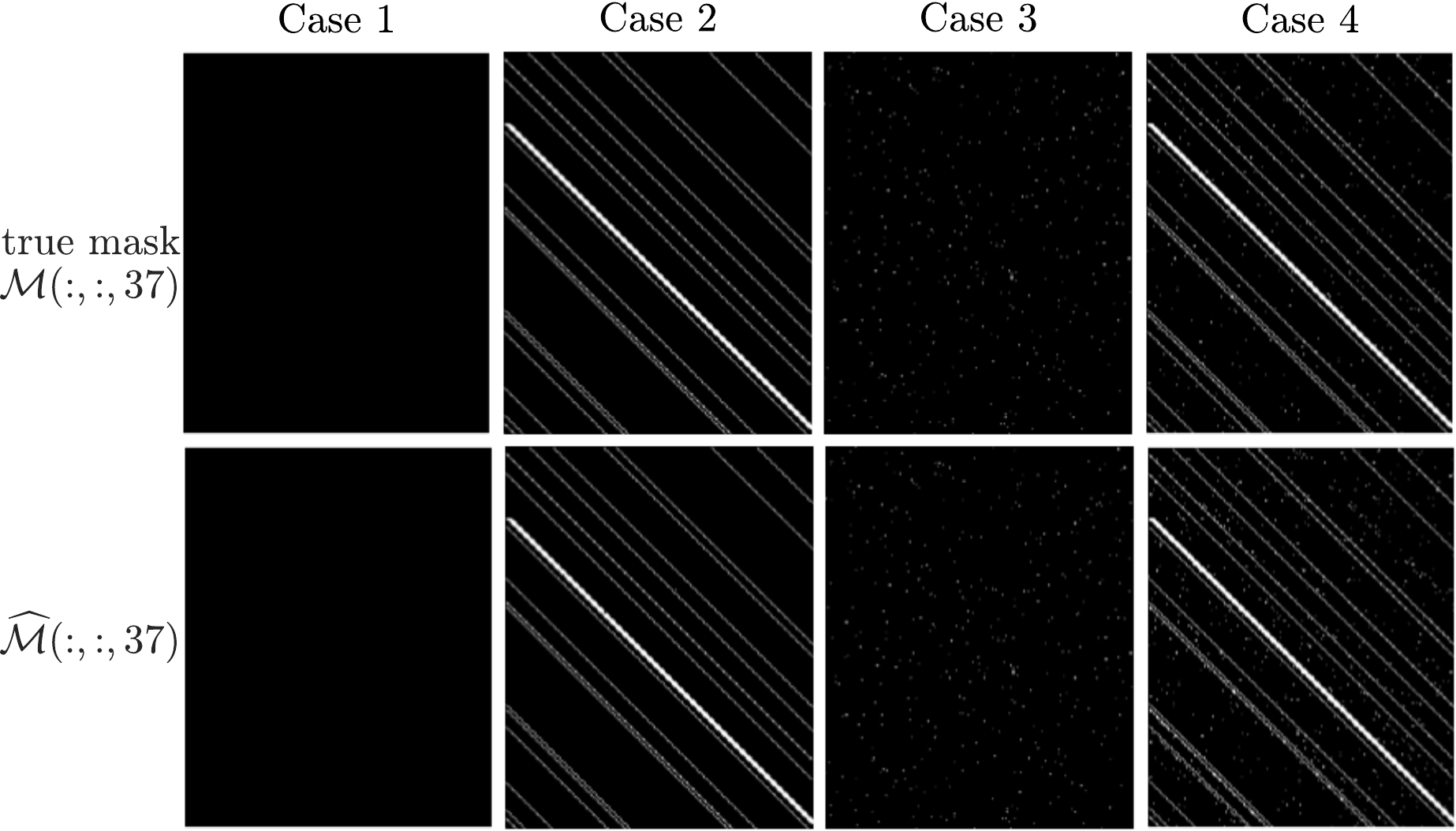}
\caption{True masks (in the first row) and its estimates   (in the second row) yielded by FastHyMix in the band 37  of simulated Pavia University  dataset.  }
\label{fig:mask}
\end{figure}

One of the contributions in this paper is to introduce
a noise estimation method elaborated for mixed noise by exploiting high spectral correlation of HSIs. The mixed noise is fitted by a Gaussian mixture model with 2 components.
The fitting enables us to make a good estimation of Gaussian noise intensity per band.

 We conducted experiments using simulated Pavia University dataset  to compare the proposed noise estimation method with a typical noise estimation method, HySime \cite{hysime}, which has been used widely in state-of-the-art denoisers \cite{8760524,NAILRMA}.
A comparison of estimated standard deviations, $\hat{\sigma}_i$, of Gaussian noise per band is  presented in Fig. \ref{fig:std_Gaussian_comparison}, where we can see both methods yield a  good estimate of $\hat{\sigma}_i$ in   case 1, where the image is corrupted only by Gaussian noise. But in cases 2-4 with mixed noise, compared with the HySime, the estimates obtained by the proposed method are much closer to the true ones. 
Our method can provide a better estimation of Gaussian noise intensity under the circumstance of mixed noise. 

The Gaussian mixture model fitted to the mixed noise also helps to identify the locations of sparse noise, represented by the mask, $\mathcal{M}$. Fig. \ref{fig:mask} displays the mask estimated by FastHyMix in band 37 of the Pavia University dataset. We can see FastHyMix can accurately identify pixels corrupted by sparse noise in all cases.

  \subsection{Deep prior for eigen-images}
   \begin{figure}[htbp]
\centering
\includegraphics[width=8cm]{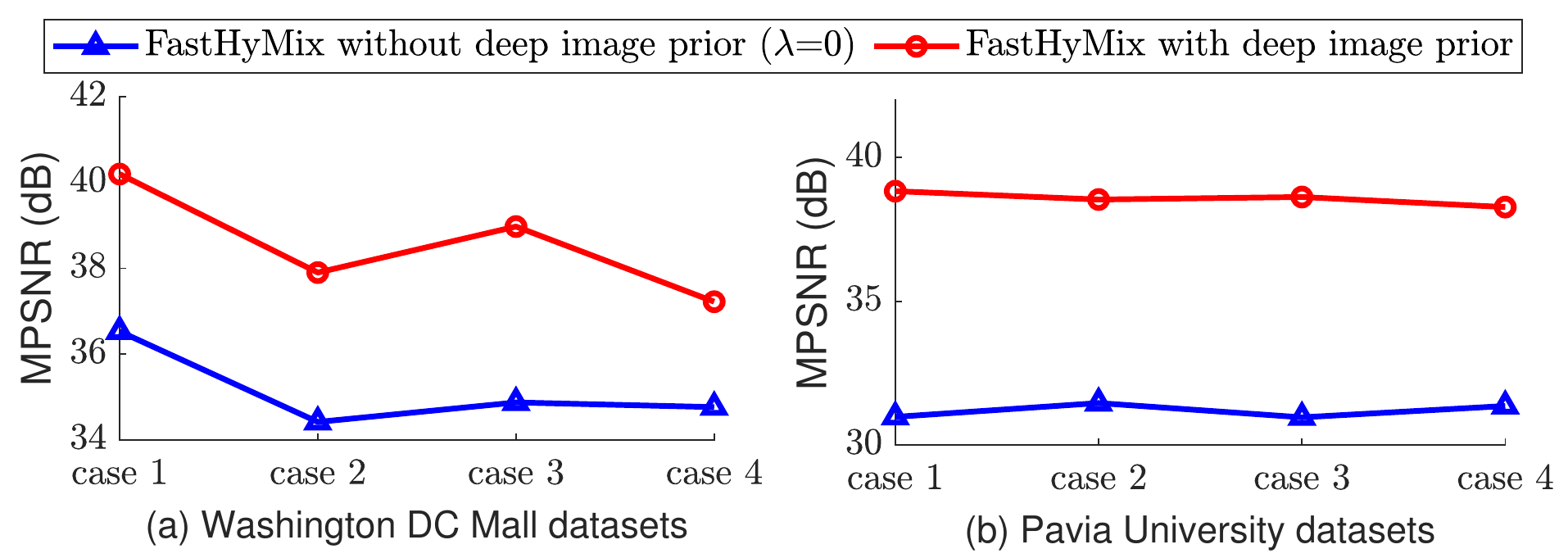}
\caption{MPSNR values of images denoised by original FastHyMix (with deep image prior) and modified FastHyMix (without deep image prior).}
 \label{fig:FastHyMix_without_deep_prior}
\end{figure}

One may question whether the deep prior from the deep network, FFDNet, which has been trained  using grayscale images acquired from commercial cameras, is suitable for remote sensing images.
To see the impact of deep prior embedded in FastHyMix, a comparison experiment using simulated datasets was conducted. Fig. \ref{fig:FastHyMix_without_deep_prior} gives MPSNR values of images denoised by original FastHyMix (with deep image prior) and modified FastHyMix (without deep image prior). MPSNR value per case is increasing considerably when eigen-images are filtered by a powerful  network, FFDNet. It demonstrates that given a good estimate of Gaussian noise level, a deep image prior make a positive contribution to  hyperspectral image denoising.  
 Although the network was trained using grayscale images, but not remote sensing images, the network still achieves impressive performance for HSI denoising. The reason is that both kinds of images are natural images, sharing same properties, such as local and non-local similarity, and piece-wise smoothness. Therefore, the image prior learned by the network from grayscale images are also applicable to HSIs.

 \subsection{Robustness of FastHyMix to subspace overestimation}
 \label{sec:robustness}

\begin{figure}[htbp]
\centering
\includegraphics[width=8cm]{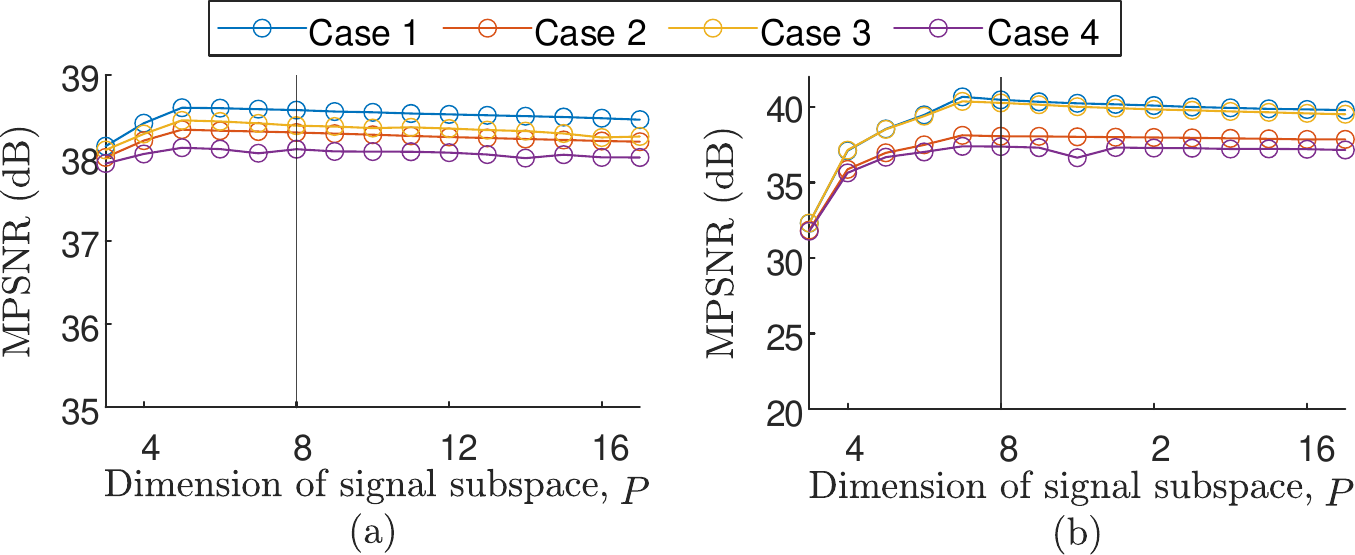}
\caption{MPSNR of FastHyMix results as a function of the dimension of the subspace estimation  (the true subspace dimension is 8) in (a) Washington DC Mall data, and (b) Pavia University data.}
 \label{fig:impactOfSubspaceDimension}
\end{figure}

The subspace dimension of two datasets input
to compared methods, NG-meet, LRTF-DFR, and FastHyMix,  was set to  8, the true
value, for fair comparison. In fact, it is challenging to estimate the subspace dimension with high accuracy  from observed HSIs corrupted by mixed noise. Fortunately, FastHyMix is robust to subspace overestimation. We take two datasets, namely, the Washington DC Mall data and Pavia University data, to show its robustness.  Fig. \ref{fig:impactOfSubspaceDimension} shows the MPSNR yielded by FastHyMix  as a function of the dimension of the subspace estimation.  
It is clear that the MPSNR is nearly constant provided that the subspace dimension is not underestimated.
To give an insight into the robustness of FastHyMix with respect to subspace overestimation, we analyse the impact of subspace overestimation on the processing of image signal, Gaussian noise, and sparse noise. 
In the objective function of FastHyMix,  \eqref{eq:opt},
as the  dimension of subspace increases, the new subspace still can represent the image well, 
but includes more amount of Gaussian noise (which  will be removed by FastHyMix via \eqref{eq:solveZ}). 
Due to the mask, $\widehat{\cal M}$,  observations with sparse noise does not contribute to the  estimate of $\widetilde{\cal Z}$ in the objective function \eqref{eq:opt}. Therefore, FastHyMix is robust to subspace dimension overestimation.

\subsection{Robustness of FastHyMix to different noise intensities and noise types}
 
\begin{figure*}[htbp]
\centering
\includegraphics[width=18cm]{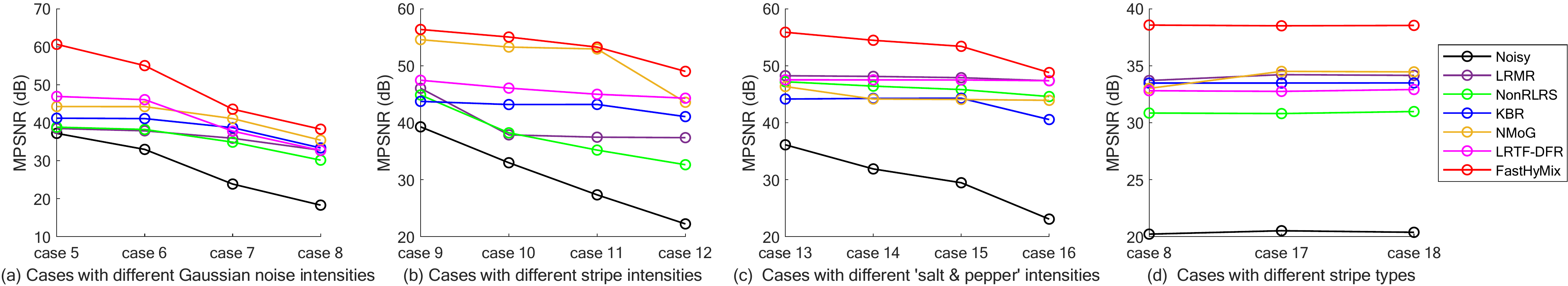}
\caption{MPSNR values of    comparison methods on Pavia University data corrupted by noise with different intensities. Noise intensity per case can be found in Tab. \ref{tab:dif_noise}.}
\label{fig:MPSNR_dif_noise_intensity}
\end{figure*}

To evaluate whether the performance of FastHyMix is robust under high-intensity or low-intensity noise
conditions, we simulated 12 kinds of mixed noise (see cases 5-16 in Tab. \ref{tab:dif_noise}) and added them into Pavia University data. 
For example, as descripbed in Tab. \ref{tab:dif_noise}, the image in case 5 is corrupted by Gaussian noise and stripes. Standard deviation of the Gaussian noise per band is sampled from a uniform distribution $U(0, 0.01)$  and oblique stripe noise randomly affected 30\% of the bands. Images in cases 5-8, in cases 9-12, and in cases 13-16 were designed to evaluate the effect of Gaussian noise intensity, stripe noise intensity, and `salt and pepper' intensity, respectively.
Denoising performance of the proposed FastHyMix and comparison methods in terms of MPSNR is shown in 
Fig. \ref{fig:MPSNR_dif_noise_intensity}-(a-c), where we can see FastHyMix uniformly yields best results in all cases, implying that FastHyMix is robust when addressing  high-intensity and low-intensity noise.

The effect of stripe value is   studied as well. Pixel value of stripes can be  the possible maximum value, the possible minimum value, or random value. We simulated above 3 kinds of stripes. The image in case 8 contains maximum-valued stripes. We generated two new noisy images similar to the image in case 8 but with minimum-valued (case 17) and random-valued stripes (case 18), respectively.  Results in Fig. \ref{fig:MPSNR_dif_noise_intensity}-(d) show the  superiority of FastHyMix over other denoising methods in cases with different stripes.

\begin{table*}[htbp]
  \centering
  \caption{A series of mixed noise with different intensities added in Pavia University data. }
    \begin{tabular}{lcccc}
    \toprule
          & Gaussian noise & \multicolumn{2}{c}{Stripe noise} & `Salt \& Pepper' noise \\
            \midrule
          & \multicolumn{1}{l}{Distribution of stardard deviation } & \multicolumn{1}{l}{Proportion of bands } & \multicolumn{1}{l}{\multirow{2}[1]{*}{Stripe value}} & \multicolumn{1}{l}{Proportion of elements in $\mathcal{X}$} \\
          & \multicolumn{1}{l}{of Gaussian noise over bands} & \multicolumn{1}{l}{affected by stripe noise} &       & \multicolumn{1}{l}{ affected by `Salt \& Pepper' } \\
    \midrule
    Case 5 & U(0, 0.01) & \multirow{4}[2]{*}{30\%} & \multirow{4}[2]{*}{maximum-valued stripes} & \multirow{4}[2]{*}{\XSolidBrush} \\
    Case 6 & U(0, 0.02) &       &       &  \\
    Case 7 & U(0.01, 0.06) &       &       &  \\
    Case 8 & U(0.05, 0.10) &       &       &  \\
    \midrule
    Case 9 & \multirow{4}[2]{*}{U(0.01, 0.06)} & 5\%   & \multirow{4}[2]{*}{maximum-valued stripes} & \multirow{4}[2]{*}{\XSolidBrush} \\
    Case 10 &       & 30\%  &       &  \\
    Case 11 &       & 50\%  &       &  \\
    Case 12 &       & 70\%  &       &  \\
    \midrule
    Case 13 & \multirow{4}[2]{*}{U(0.01, 0.06)} & \multirow{4}[2]{*}{\XSolidBrush} & \multirow{4}[2]{*}{maximum-valued stripes} & 0.01\% \\
    Case 14 &       &       &       & 0.05\% \\
    Case 15 &       &       &       & 0.10\% \\
    Case 16 &       &       &       & 0.50\% \\
    \midrule
    Case 17 & \multirow{2}[2]{*}{U(0.05, 0.10)} & \multirow{2}[2]{*}{30\%} & minimum-valued stripes & \multirow{2}[2]{*}{\XSolidBrush} \\
    Case 18 &       &       & random-valued stripes &  \\
    \bottomrule
    \end{tabular}%
  \label{tab:dif_noise}%
\end{table*}%

\section{Experiments with real images}
\label{sec:exp_real}
The performance of hyperspectral mixed noise removal methods is also tested 
on two real HSI datasets, namely Tiangong-1 image and Hyperion Cuprite image, shown in Fig. \ref{fig:RealImg}-(c,d).

\subsection{Tiangong-1 dataset}

\begin{figure*}[htbp]
\centering
\includegraphics[width=16cm]{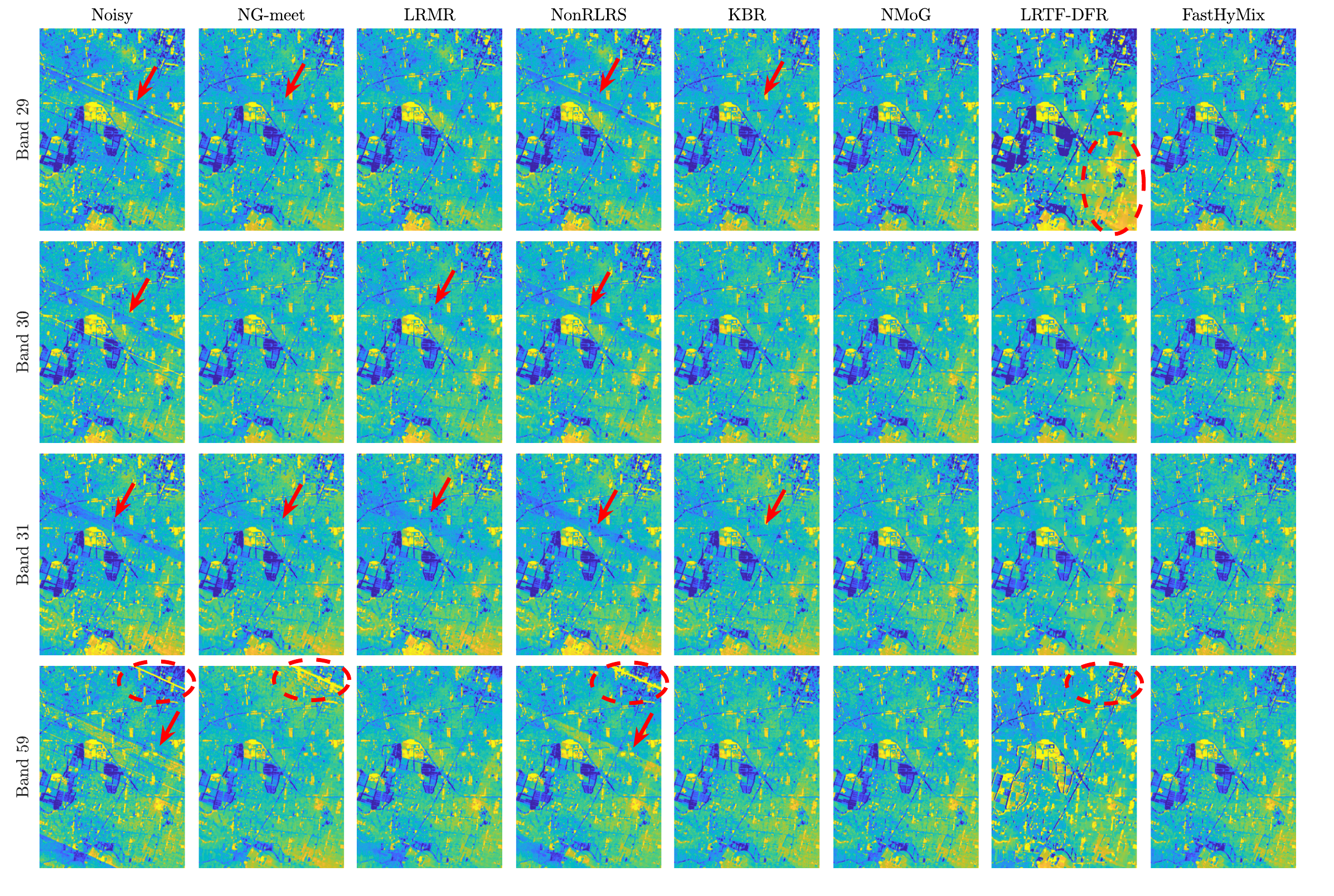}
\caption{Recovered bands obtained by NG-meet (63 s), LRMR (30 s), NonRLRS (29 s), KBR (203 s), NMoG (90 s), LRTF-DFR (85 s), and FastHyMix (7 s) in the real HSI Tiangong-1
data. Red arrows and dash circles are added to mark the artifacts. }
\label{fig:TG_bands}
\end{figure*}

The Tiangong-1 dataset  was acquired over an area of Qinghai Province, China in May 2013, by a sensor placed in Tiangong-1  imager, which has a 75-band push broom scanner with nominal bandwidth of 23 nm short wave infrared (SWIR), covering from 800nm to 2500 nm.
 A subregion image of size $351 \times 253$ pixels was tested.
Five bands displaying strong noise are shown in Fig.  \ref{fig:TG_bands}, where bands 29, 30, 31, and 59 contain obvious stripes. Comparing the images before denoising and after denoising, we can see that   NMoG, LRTF-DFR, and FastHyMix can alleviate stripe noise in these bands.  
If we focus on   bands 29 and 59, LRTF-DFR obtains results with incorrect illumination (see area marked by red circles). 
Computational time of each method is reported in the
figure caption. Qualitatively, FastHyMix yields the best result in
the shortest time.

\subsection{Hyperion Cuprite dataset}

\begin{figure*}[htbp]
\centering
\includegraphics[width=16cm]{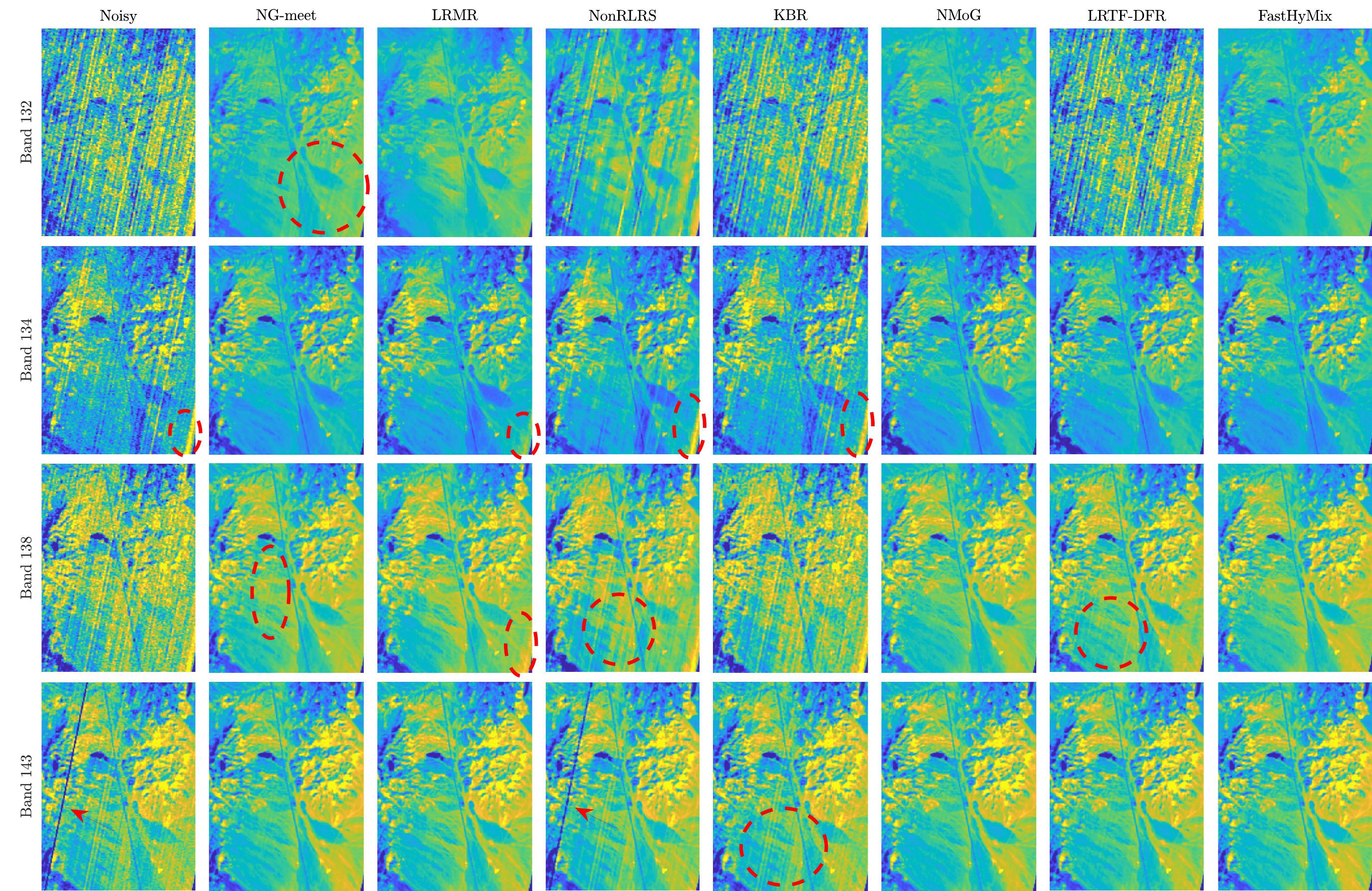}
\caption{ Recovered bands obtained by NG-meet (45 s), LRMR (28 s), NonRLRS (38 s), KBR (248 s), NMoG (83 s), LRTF-DFR (58 s), and FastHyMix (7 s) in the real HSI Hyperion Cuprite data. Red arrows and dash circles are added to mark the artifacts.}
\label{fig:Cuprite_bands}
\end{figure*}

The Cuprite HSI  was captured at Cuprite, NV, USA, by Hyperion   sensor, which divides the spectrum from 355  nm to 2577  nm into 242 channels with a spectral resolution of 10 nm. Spatial resolution of the image is 30 meters. A subregion of size $240 \times 178$ pixels  with  177 spectral channels (after removing water vapour absorption bands)  is cropped for test. Four   bands  are shown in Fig. \ref{fig:Cuprite_bands}, where bands 132, 134, and 138 are corrupted by severe stripes, and band    143  is mainly affected by   deadlines.
All methods (except NonRLRS) are able to remove the deadlines in bands 130 and 143. For severe stripes in the other three bands,
 we can see  that   NMoG  and FastHyMix 
 achieved good restoration results while
obvious stripes remained within the results of other methods.
 As shown in the figure caption, the running time of FastHyMix is much shorter than the comparison methods.

\section{Conclusion}
\label{sec:concl}

 This paper introduce a fast and user-friendly hyperspectral mixed noise removal method, FastHyMix. 
It fits the mixed noise using a Gaussian mixture model, which is a universal approximation to any continuous distribution and hence
capable of modelling a complex noise distribution. The fitting model enables us to make a good estimation of Gaussian noise intensity per band and the locations  of sparse noise.  The characteristics of HSIs, namely, spectral low-rankness and high spatial correlation, are exploited by using a subspace representation and a deep image prior. The proposed method has some merits: a)
FastHyMix method is user-friendly in the sense that its regularization parameters are set adaptively to the noise statistics.
b) A comparison of FastHyMix with the state-of-the-art algorithms was conducted,
leading to the conclusion that FastHyMix yields similar
or better performance for complex mixed noise, with much shorter running time.
These characteristics put FastHyMix in a
superior position to be used as an HSI denoiser.

\ifCLASSOPTIONcaptionsoff
  \newpage
\fi

\bibliographystyle{IEEEtran}
\bibliography{FastHyMix}

\begin{thebibliography}{10}
\providecommand{\url}[1]{#1}
\csname url@samestyle\endcsname
\providecommand{\newblock}{\relax}
\providecommand{\bibinfo}[2]{#2}
\providecommand{\BIBentrySTDinterwordspacing}{\spaceskip=0pt\relax}
\providecommand{\BIBentryALTinterwordstretchfactor}{4}
\providecommand{\BIBentryALTinterwordspacing}{\spaceskip=\fontdimen2\font plus
\BIBentryALTinterwordstretchfactor\fontdimen3\font minus
  \fontdimen4\font\relax}
\providecommand{\BIBforeignlanguage}[2]{{%
\expandafter\ifx\csname l@#1\endcsname\relax
\typeout{** WARNING: IEEEtran.bst: No hyphenation pattern has been}%
\typeout{** loaded for the language `#1'. Using the pattern for}%
\typeout{** the default language instead.}%
\else
\language=\csname l@#1\endcsname
\fi
#2}}
\providecommand{\BIBdecl}{\relax}
\BIBdecl

\bibitem{overview}
J.~Bioucas-Dias, A.~Plaza, N.~Dobigeon, M.~Parente, Q.~Du, P.~Gader, and
  J.~Chanussot, ``Hyperspectral unmixing overview: Geometrical, statistical,
  and sparse regression-based approaches,'' \emph{IEEE Journal of Selected
  Topics in Applied Earth Observations and Remote Sensing}, vol.~5, no.~2, pp.
  354--379, Apr. 2012.

\bibitem{8295275}
R.~Dian, S.~Li, A.~Guo, and L.~Fang, ``Deep hyperspectral image sharpening,''
  \emph{IEEE Transactions on Neural Networks and Learning Systems}, vol.~29,
  no.~11, pp. 5345--5355, 2018.

\bibitem{8750899}
W.~Xie, J.~Lei, Y.~Cui, Y.~Li, and Q.~Du, ``Hyperspectral pansharpening with
  deep priors,'' \emph{IEEE Transactions on Neural Networks and Learning
  Systems}, vol.~31, no.~5, pp. 1529--1543, 2020.

\bibitem{9136736}
L.~Zhang, J.~Nie, W.~Wei, Y.~Li, and Y.~Zhang, ``Deep blind hyperspectral image
  super-resolution,'' \emph{IEEE Transactions on Neural Networks and Learning
  Systems}, pp. 1--13, 2020.

\bibitem{9130919}
C.-H. Lin and J.~M. Bioucas-Dias, ``Nonnegative blind source separation for
  ill-conditioned mixtures via john ellipsoid,'' \emph{IEEE Transactions on
  Neural Networks and Learning Systems}, vol.~32, no.~5, pp. 2209--2223, 2021.

\bibitem{9404853}
J.~Liu, Z.~Hou, W.~Li, R.~Tao, D.~Orlando, and H.~Li, ``Multipixel anomaly
  detection with unknown patterns for hyperspectral imagery,'' \emph{IEEE
  Transactions on Neural Networks and Learning Systems}, pp. 1--11, 2021.

\bibitem{9288702}
L.~Li, W.~Li, Y.~Qu, C.~Zhao, R.~Tao, and Q.~Du, ``Prior-based tensor
  approximation for anomaly detection in hyperspectral imagery,'' \emph{IEEE
  Transactions on Neural Networks and Learning Systems}, pp. 1--14, 2020.

\bibitem{FastHyDe}
L.~Zhuang and J.~Bioucas-Dias, ``Fast hyperspectral image denoising and
  inpainting based on low-rank and sparse representations,'' \emph{IEEE Journal
  of Selected Topics in Applied Earth Observations and Remote Sensing},
  vol.~11, no.~3, pp. 730--742, 2018.

\bibitem{chang2013simultaneous}
Y.~Chang, L.~Yan, H.~Fang, and H.~Liu, ``Simultaneous destriping and denoising
  for remote sensing images with unidirectional total variation and sparse
  representation,'' \emph{IEEE Geoscience and Remote Sensing Letters}, vol.~11,
  no.~6, pp. 1051--1055, 2013.

\bibitem{chang2017hyper}
Y.~Chang, L.~Yan, and S.~Zhong, ``Hyper-laplacian regularized unidirectional
  low-rank tensor recovery for multispectral image denoising,'' in
  \emph{Proceedings of the IEEE Conference on Computer Vision and Pattern
  Recognition}, 2017, pp. 4260--4268.

\bibitem{jiang2007hybrid}
S.~Jiang and X.~Hao, ``Hybrid fourier-wavelet image denoising,''
  \emph{Electronics Letters}, vol.~43, no.~20, pp. 1081--1082, 2007.

\bibitem{rasti2013hyperspectral}
B.~Rasti, J.~R. Sveinsson, M.~O. Ulfarsson, and J.~A. Benediktsson,
  ``Hyperspectral image denoising using first order spectral roughness penalty
  in wavelet domain,'' \emph{IEEE Journal of Selected Topics in Applied Earth
  Observations and Remote Sensing}, vol.~7, no.~6, pp. 2458--2467, 2013.

\bibitem{NMoG}
Y.~Chen, X.~Cao, Q.~Zhao, D.~Meng, and Z.~Xu, ``Denoising hyperspectral image
  with non-i.i.d. noise structure,'' \emph{IEEE Transactions on Cybernetics},
  vol.~48, no.~3, pp. 1054--1066, 2018.

\bibitem{LRTF-DFR}
Y.-B. Zheng, T.-Z. Huang, X.-L. Zhao, Y.~Chen, and W.~He,
  ``Double-factor-regularized low-rank tensor factorization for mixed noise
  removal in hyperspectral image,'' \emph{IEEE Transactions on Geoscience and
  Remote Sensing}, vol.~58, no.~12, pp. 8450--8464, 2020.

\bibitem{he2019non}
W.~He, Q.~Yao, C.~Li, N.~Yokoya, Q.~Zhao, H.~Zhang, and L.~Zhang, ``Non-local
  meets global: An integrated paradigm for hyperspectral image restoration,''
  \emph{IEEE Transactions on Pattern Analysis and Machine Intelligence}, pp.
  1--1, 2020.

\bibitem{L1HyMixDe}
L.~{Zhuang} and M.~K. {Ng}, ``Hyperspectral mixed noise removal by $\ell
  _1$-norm-based subspace representation,'' \emph{IEEE Journal of Selected
  Topics in Applied Earth Observations and Remote Sensing}, vol.~13, pp.
  1143--1157, 2020.

\bibitem{LRMR}
H.~{Zhang}, W.~{He}, L.~{Zhang}, H.~{Shen}, and Q.~{Yuan}, ``Hyperspectral
  image restoration using low-rank matrix recovery,'' \emph{IEEE Transactions
  on Geoscience and Remote Sensing}, vol.~52, no.~8, pp. 4729--4743, 2014.

\bibitem{8760524}
T.~{Xie}, S.~{Li}, and B.~{Sun}, ``Hyperspectral images denoising via nonconvex
  regularized low-rank and sparse matrix decomposition,'' \emph{IEEE
  Transactions on Image Processing}, pp. 1--1, Jul. 2019.

\bibitem{9374571}
H.~Zhang, J.~Cai, W.~He, H.~Shen, and L.~Zhang, ``Double low-rank matrix
  decomposition for hyperspectral image denoising and destriping,'' \emph{IEEE
  Transactions on Geoscience and Remote Sensing}, pp. 1--19, 2021.

\bibitem{KBR}
Q.~Xie, Q.~Zhao, D.~Meng, and Z.~Xu, ``Kronecker-basis-representation based
  tensor sparsity and its applications to tensor recovery,'' \emph{IEEE
  Transactions on Pattern Analysis and Machine Intelligence}, vol.~40, no.~8,
  pp. 1888--1902, 2018.

\bibitem{SSAHTV}
Q.~Yuan, L.~Zhang, and H.~Shen, ``Hyperspectral image denoising employing a
  spectral--spatial adaptive total variation model,'' \emph{IEEE Transactions
  on Geoscience and Remote Sensing}, vol.~50, no.~10, pp. 3660--3677, 2012.

\bibitem{zhao2014hyperspectral}
Y.-Q. Zhao and J.~Yang, ``Hyperspectral image denoising via sparse
  representation and low-rank constraint,'' \emph{IEEE Transactions on
  Geoscience and Remote Sensing}, vol.~53, no.~1, pp. 296--308, 2014.

\bibitem{dian_8359412}
S.~Li, R.~Dian, L.~Fang, and J.~M. Bioucas-Dias, ``Fusing hyperspectral and
  multispectral images via coupled sparse tensor factorization,'' \emph{IEEE
  Transactions on Image Processing}, vol.~27, no.~8, pp. 4118--4130, 2018.

\bibitem{zheng2019mixed}
Y.-B. Zheng, T.-Z. Huang, X.-L. Zhao, T.-X. Jiang, T.-H. Ma, and T.-Y. Ji,
  ``Mixed noise removal in hyperspectral image via low-fibered-rank
  regularization,'' \emph{IEEE Transactions on Geoscience and Remote Sensing},
  vol.~58, no.~1, pp. 734--749, 2019.

\bibitem{LIN2021126342}
\BIBentryALTinterwordspacing
J.~Lin, T.-Z. Huang, X.-L. Zhao, T.-H. Ma, T.-X. Jiang, and Y.-B. Zheng, ``A
  novel non-convex low-rank tensor approximation model for hyperspectral image
  restoration,'' \emph{Applied Mathematics and Computation}, vol. 408, p.
  126342, 2021. [Online]. Available:
  \url{https://www.sciencedirect.com/science/article/pii/S0096300321004318}
\BIBentrySTDinterwordspacing

\bibitem{TenSR}
J.~Lin, T.-Z. Huang, X.-L. Zhao, T.-X. Jiang, and L.~Zhuang, ``A tensor
  subspace representation-based method for hyperspectral image denoising,''
  \emph{IEEE Transactions on Geoscience and Remote Sensing}, pp. 1--19, 2020.

\bibitem{chang2016remote}
Y.~Chang, L.~Yan, T.~Wu, and S.~Zhong, ``Remote sensing image stripe noise
  removal: From image decomposition perspective,'' \emph{IEEE Transactions on
  Geoscience and Remote Sensing}, vol.~54, no.~12, pp. 7018--7031, 2016.

\bibitem{gao2014subspace}
L.~Gao, J.~Li, M.~Khodadadzadeh, A.~Plaza, B.~Zhang, Z.~He, and H.~Yan,
  ``Subspace-based support vector machines for hyperspectral image
  classification,'' \emph{IEEE Geoscience and Remote Sensing Letters}, vol.~12,
  no.~2, pp. 349--353, 2014.

\bibitem{AdeHyDe}
T.~{Jiang}, L.~{Zhuang}, T.~{Huang}, and J.~M. {Bioucas-Dias}, ``Adaptive
  hyperspectral mixed noise removal,'' in \emph{IEEE International Geoscience
  and Remote Sensing Symposium}, Jul. 2018, pp. 4035--4038.

\bibitem{8718504}
R.~{Dian} and S.~{Li}, ``Hyperspectral image super-resolution via
  subspace-based low tensor multi-rank regularization,'' \emph{IEEE
  Transactions on Image Processing}, pp. 1--1, 2019.

\bibitem{Dian2019}
R.~Dian, S.~Li, and L.~Fang, ``Learning a low tensor-train rank representation
  for hyperspectral image super-resolution,'' \emph{IEEE Transactions on Neural
  Networks and Learning Systems}, vol.~30, no.~9, pp. 2672--2683, 2019.

\bibitem{8948303}
Y.~Xu, Z.~Wu, J.~Chanussot, and Z.~Wei, ``Hyperspectral images super-resolution
  via learning high-order coupled tensor ring representation,'' \emph{IEEE
  Transactions on Neural Networks and Learning Systems}, vol.~31, no.~11, pp.
  4747--4760, 2020.

\bibitem{dian2020recent}
R.~Dian, S.~Li, B.~Sun, and A.~Guo, ``Recent advances and new guidelines on
  hyperspectral and multispectral image fusion,'' \emph{Information Fusion},
  2020.

\bibitem{dian2019nonlocal}
R.~Dian, S.~Li, L.~Fang, T.~Lu, and J.~M. Bioucas-Dias, ``Nonlocal sparse
  tensor factorization for semiblind hyperspectral and multispectral image
  fusion,'' \emph{IEEE transactions on cybernetics}, vol.~50, no.~10, pp.
  4469--4480, 2019.

\bibitem{GLF}
L.~Zhuang, X.~Fu, M.~K. Ng, and J.~M. Bioucas-Dias, ``Hyperspectral image
  denoising based on global and nonlocal low-rank factorizations,'' \emph{IEEE
  Transactions on Geoscience and Remote Sensing}, 2021.

\bibitem{cao2019hyperspectral}
C.~Cao, J.~Yu, C.~Zhou, K.~Hu, F.~Xiao, and X.~Gao, ``Hyperspectral image
  denoising via subspace-based nonlocal low-rank and sparse factorization,''
  \emph{IEEE Journal of Selected Topics in Applied Earth Observations and
  Remote Sensing}, vol.~12, no.~3, pp. 973--988, 2019.

\bibitem{BM3D}
K.~Dabov, A.~Foi, V.~Katkovnik, and K.~Egiazarian, ``Image denoising by sparse
  {3-D} transform-domain collaborative filtering,'' \emph{IEEE Transactions on
  Image Processing}, vol.~16, no.~8, pp. 2080--2095, Aug. 2007.

\bibitem{WNNM}
S.~Gu, L.~Zhang, W.~Zuo, and X.~Feng, ``Weighted nuclear norm minimization with
  application to image denoising,'' in \emph{Proceedings of the IEEE conference
  on computer vision and pattern recognition}, 2014, pp. 2862--2869.

\bibitem{DnCNN}
K.~{Zhang}, W.~{Zuo}, Y.~{Chen}, D.~{Meng}, and L.~{Zhang}, ``Beyond a gaussian
  denoiser: Residual learning of deep cnn for image denoising,'' \emph{IEEE
  Transactions on Image Processing}, vol.~26, no.~7, pp. 3142--3155, 2017.

\bibitem{FFDNet}
K.~{Zhang}, W.~{Zuo}, and L.~{Zhang}, ``{FFDNet}: Toward a fast and flexible
  solution for cnn-based image denoising,'' \emph{IEEE Transactions on Image
  Processing}, vol.~27, no.~9, pp. 4608--4622, 2018.

\bibitem{CBDNet}
S.~{Guo}, Z.~{Yan}, K.~{Zhang}, W.~{Zuo}, and L.~{Zhang}, ``Toward
  convolutional blind denoising of real photographs,'' in \emph{IEEE/CVF
  Conference on Computer Vision and Pattern Recognition (CVPR)}, 2019, pp.
  1712--1722.

\bibitem{SSGN}
Q.~Zhang, Q.~Yuan, J.~Li, X.~Liu, H.~Shen, and L.~Zhang, ``{Hybrid noise
  removal in hyperspectral imagery with a spatial-spectral gradient network},''
  \emph{IEEE Transactions on Geoscience and Remote Sensing}, vol.~57, no.~10,
  pp. 7317--7329, 2019.

\bibitem{HSI-DeNet}
Y.~Chang, L.~Yan, H.~Fang, S.~Zhong, and W.~Liao, ``{HSI-DeNet: Hyperspectral
  Image Restoration via Convolutional Neural Network},'' \emph{IEEE
  Transactions on Geoscience and Remote Sensing}, vol.~57, no.~2, pp. 667--682,
  2019.

\bibitem{DSSBP}
Q.~Zhang, Q.~Yuan, J.~Li, F.~Sun, and L.~Zhang, ``{Deep spatio-spectral
  Bayesian posterior for hyperspectral image non-i.i.d. noise removal},''
  \emph{ISPRS Journal of Photogrammetry and Remote Sensing}, vol. 164, no.
  November 2019, pp. 125--137, 2020.

\bibitem{3DADNet}
Q.~Shi, X.~Tang, T.~Yang, R.~Liu, and L.~Zhang, ``Hyperspectral image denoising
  using a 3-d attention denoising network,'' \emph{IEEE Transactions on
  Geoscience and Remote Sensing}, pp. 1--16, 2021.

\bibitem{6737048}
S.~V. {Venkatakrishnan}, C.~A. {Bouman}, and B.~{Wohlberg}, ``Plug-and-play
  priors for model based reconstruction,'' in \emph{2013 IEEE Global Conference
  on Signal and Information Processing}, Dec. 2013, pp. 945--948.

\bibitem{chan2016plug}
S.~H. Chan, X.~Wang, and O.~A. Elgendy, ``Plug-and-play admm for image
  restoration: Fixed-point convergence and applications,'' \emph{IEEE
  Transactions on Computational Imaging}, vol.~3, no.~1, pp. 84--98, 2016.

\bibitem{dian2020regularizing}
R.~Dian, S.~Li, and X.~Kang, ``Regularizing hyperspectral and multispectral
  image fusion by cnn denoiser,'' \emph{IEEE transactions on neural networks
  and learning systems}, 2020.

\bibitem{romano2017little}
Y.~Romano, M.~Elad, and P.~Milanfar, ``The little engine that could:
  Regularization by denoising (red),'' \emph{SIAM Journal on Imaging Sciences},
  vol.~10, no.~4, pp. 1804--1844, 2017.

\bibitem{8305626}
X.~Chen, Z.~Han, Y.~Wang, Q.~Zhao, D.~Meng, L.~Lin, and Y.~Tang, ``A
  generalized model for robust tensor factorization with noise modeling by
  mixture of gaussians,'' \emph{IEEE Transactions on Neural Networks and
  Learning Systems}, vol.~29, no.~11, pp. 5380--5393, 2018.

\bibitem{hysime}
J.~M. {Bioucas-Dias} and J.~M.~P. {Nascimento}, ``Hyperspectral subspace
  identification,'' \emph{IEEE Transactions on Geoscience and Remote Sensing},
  vol.~46, no.~8, pp. 2435--2445, Aug. 2008.

\bibitem{gao2013comparative}
L.~Gao, Q.~Du, B.~Zhang, W.~Yang, and Y.~Wu, ``A comparative study on linear
  regression-based noise estimation for hyperspectral imagery,'' \emph{IEEE
  Journal of Selected Topics in Applied Earth Observations and Remote Sensing},
  vol.~6, no.~2, pp. 488--498, 2013.

\bibitem{bai2005tests}
J.~Bai and S.~Ng, ``Tests for skewness, kurtosis, and normality for time series
  data,'' \emph{Journal of Business \& Economic Statistics}, vol.~23, no.~1,
  pp. 49--60, 2005.

\bibitem{EM}
G.~J. McLachlan, S.~X. Lee, and S.~I. Rathnayake, ``{Finite mixture models},''
  \emph{Annual Review of Statistics and Its Application}, vol.~6, no. 1988, pp.
  355--378, 2019.

\bibitem{RhyDe}
L.~{Zhuang}, L.~{Gao}, B.~{Zhang}, X.~{Fu}, and J.~M. {Bioucas-Dias},
  ``Hyperspectral image denoising and anomaly detection based on low-rank and
  sparse representations,'' \emph{IEEE Transactions on Geoscience and Remote
  Sensing}, pp. 1--17, 2020.

\bibitem{Fu2021}
X.~{Fu}, S.~{Jia}, L.~{Zhuang}, M.~{Xu}, J.~{Zhou}, and Q.~{Li},
  ``Hyperspectral anomaly detection via deep plug-and-play denoising {CNN}
  regularization,'' \emph{IEEE Transactions on Geoscience and Remote Sensing},
  pp. 1--16, 2021.

\bibitem{jain2008natural}
V.~Jain and S.~Seung, ``Natural image denoising with convolutional networks,''
  \emph{Advances in neural information processing systems}, vol.~21, pp.
  769--776, 2008.

\bibitem{SSIM}
Z.~Wang, A.~C. Bovik, H.~R. Sheikh, and E.~P. Simoncelli, ``Image quality
  assessment: from error visibility to structural similarity,'' \emph{IEEE
  transactions on image processing}, vol.~13, no.~4, pp. 600--612, 2004.

\bibitem{FSIM}
L.~Zhang, L.~Zhang, X.~Mou, and D.~Zhang, ``{FSIM: A} feature similarity index
  for image quality assessment,'' \emph{IEEE transactions on Image Processing},
  vol.~20, no.~8, pp. 2378--2386, 2011.

\bibitem{NAILRMA}
W.~He, H.~Zhang, L.~Zhang, and H.~Shen, ``Hyperspectral image denoising via
  noise-adjusted iterative low-rank matrix approximation,'' \emph{IEEE Journal
  of Selected Topics in Applied Earth Observations and Remote Sensing}, vol.~8,
  no.~6, pp. 3050--3061, Jun. 2015.

\end{thebibliography}

\begin{IEEEbiography}
[{\includegraphics[width=1in,height=1.25in,clip,keepaspectratio]{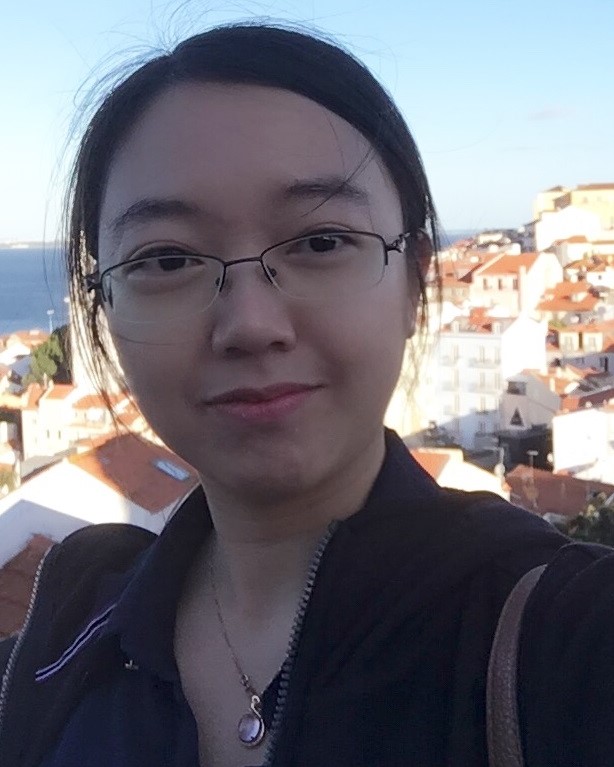}}]{Lina Zhuang}
(S'15-M'20) received Bachelor's  degrees in geographic information system and in economics from South China Normal University, Ghuangzhou, China,  in  2012,  the  M.S.  degree  in  cartography  and  geography  information  system  from  Institute  of  Remote  Sensing  and  Digital  Earth,  Chinese Academy of Sciences, Beijing, China, in 2015, and the Ph.D. degree in Electrical and Computer Engineering at the Instituto Superior Tecnico,  Universidade de Lisboa, Lisbon, Portugal in 2018.

From 2015 to 2018, she was  a Marie Curie Early Stage Researcher of Sparse Representations and Compressed Sensing Training Network (SpaRTaN number 607290) with the Instituto de Telecomunica\c{c}\~{o}es. SpaRTaN Initial Training Networks (ITN) is funded under the European Union's Seventh Framework Programme  (FP7-PEOPLE-2013-ITN)  call  and  is  part  of  the  Marie  Curie Actions-ITN  funding  scheme. From 2019 to 2021, she was a Research Assistant Professor with    Hong Kong Baptist University.  She is currently  a Research Assistant Professor with the University of Hong Kong. Her research interests include hyperspectral image restoration, superresolution, and compressive sensing.
\end{IEEEbiography}

 \begin{IEEEbiography}
[{\includegraphics[width=1in,height=1.25in,clip,keepaspectratio]{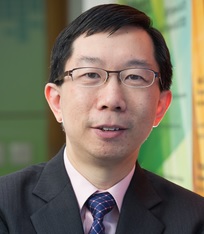}}]{Michael K. Ng}
(Senior Member, IEEE) received the B.Sc. and M.Phil. degrees from The University
of Hong Kong in 1990 and 1992, respectively, and the Ph.D. degree from The Chinese University of Hong Kong in 1995. From 1995 to 1997, he was a Research Fellow with the Computer Sciences
Laboratory, Australian National University, and an Assistant Professor/an Associate Professor with The
University of Hong Kong from 1997 to 2005. From 2006 to 2019, he was a Professor/the Chair Professor with the Department of Mathematics, Hong Kong Baptist University. He is currently a Chair Professor with the Research Division of Mathematical and Statistical Science, The University of Hong Kong. His research interests include bioinformatics, image processing, scientific computing, and data mining. He is selected for the 2017 class of fellows
of the Society for Industrial and Applied Mathematics. He received the Feng Kang Prize for his significant contributions in scientific computing. He serves as the editorial board member for several international journals.
\end{IEEEbiography}

\end{document}